\documentclass[aps,prl,superscriptaddress,twocolumn,eprint,nofootinbib]{revtex4-2}
\usepackage{amsmath}
\usepackage{amsthm}
\usepackage{mathtools}
\usepackage{hyperref}
\usepackage{amssymb}
\usepackage{physics}
\usepackage{graphicx}
\graphicspath{ {figures/} }
\usepackage[caption=false]{subfig}
\usepackage{bm}
\usepackage[dvipsnames]{xcolor}
\usepackage{float}

\usepackage{setspace} % allow double space

\usepackage[section]{placeins} % should prevent floats from being moved to different section

\usepackage{comment}

\usepackage{empheq}

\usepackage{tikz}
\usetikzlibrary{positioning, calc}

\newcommand{\p}[2]{\ensuremath{\frac{\partial #1}{\partial #2}}} %used to write partial derivatives

\newcommand{\re}[1]{\ensuremath{\text{Re}\left[ #1\right]}}

\newcommand{\beq}{\begin{equation}}
	\newcommand{\eeq}{\end{equation}}

\begin{document}
	%	\DeclareFixedFont{\eighteenpt}{\encodingdefault}{\familydefault}{\seriesdefault}{\shapedefault}{18}
	%{\eighteenpt 
		
		\title{Phase diagram, confining strings, and a new universality class in nematopolar matter}
		\author{Farzan Vafa}
		\affiliation{Center of Mathematical Sciences and Applications, Harvard University, Cambridge, MA 02138, USA}
        \affiliation{Physics of Living Systems, Department of Physics, Massachusetts Institute of Technology, Cambridge, Massachusetts 02139, USA}
		\author{Amin Doostmohammadi}
		\affiliation{Niels Bohr Institute, University of Copenhagen, Blegdamsvej 17, Copenhagen 2100, Denmark}
		% \pmb out how to write date
		\date{\today}

	\begin{abstract}

We study a minimal model of a system with coexisting nematic and polar orientational orders, where one field tends to order and the other prefers isotropy. For strong coupling, the ordered field aligns the isotropic one, locking their orientations. The phase diagram reveals three distinct phases—nematopolar (aligned orders), nematic (independent orders), and isotropic (vanishing orders)—separated by continuous and discontinuous transitions, including a triple and a tricritical point. We find unique critical scaling for the nematopolar-nematic transition, distinct from standard nematic or polar universality classes. Additionally, in the locked nematopolar phase, we show nematic $+1/2$ topological defect pairs are connected and confined by {\it strings} with constant tension. These strings arise from frustration in locking the orientational orders and can be interpreted as elongated cores of $+1$ polar topological defects. When a sufficiently strong background field couples to the polar order, all topological defects are expelled from the region. Analytical predictions are quantitatively confirmed by numerical simulations.
  
	\end{abstract}
	
	\maketitle

Orientational order, characterized by the alignment of constituent units in preferred directions, is a fundamental phenomenon observed across a wide range of natural and synthetic systems~\cite{gennes1993the}. This order manifests in various forms, from the liquid crystals in display technologies (LCDs)~\cite{yang2014fundamentals} to the collective behaviors seen in biological tissues~\cite{doostmohammadi2021physics}. Common examples include polar order, where units align in a specific direction, akin to the coordinated movement of a flock of birds~\cite{marchetti2013hydrodynamics} or the arrangement of molecules in Langmuir-Blodgett monolayers at a liquid-air interface~\cite{kaganer1999structure}. Similarly, nematic order is observed when units align along a common axis, as seen in liquid crystal sensors~\cite{schenning2017liquid}, nematic elastomers~\cite{ula2018liquid}, bacterial aggregates~\cite{volfson2008biomechanical,doostmohammadi2016defect}, suspensions of cytoskeletal filaments and associated motor proteins~\cite{kruse2005generic}, or in epithelial monolayers~\cite{saw2017topological}.
	\begin{figure}[b!]
	\includegraphics[width=0.8\columnwidth]{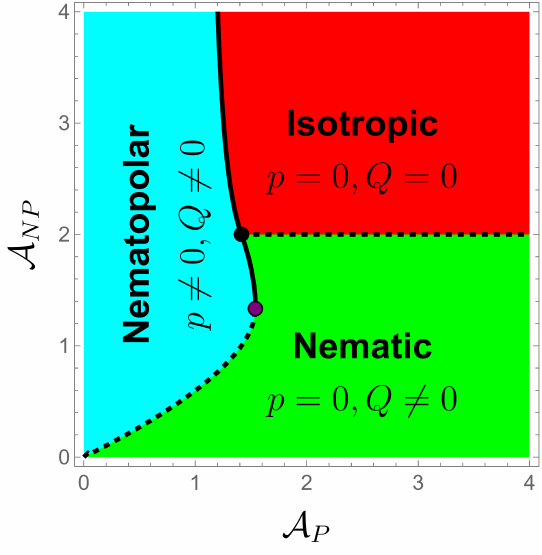}
	\caption{Phase diagram of a system with coexisting nematic and polar orders. The solid lines mark first order phase transitions, and the dashed lines are second order phase transitions. The tricritical point (purple filled circle) is at $(\mathcal A_P,\mathcal A_{NP}) = (8/\sqrt{27}, 4/3)$, and the triple point (black filled circle) is at $(\mathcal A_P,\mathcal A_{NP}) = (\sqrt{2}, 2)$. The dashed phase boundary connecting the origin to the white point satisfies Eq.~\eqref{eq:continuous}. The other dashed phase boundary is given by $\mathcal A_{NP} = 2$ for $\mathcal A_P > \sqrt{2}$. The lower and upper segments of the solid phase boundary satisfy Eqs.~\eqref{eq:discontinuous_lower} and \eqref{eq:discontinous_upper}, respectively.}
	\label{fig:phase_diagram}
\end{figure}

In many real-world materials, these types of orientational order do not exist in isolation but often coexist and interact, leading to complex phases. Recent studies have uncovered phases such as the ferroelectric nematic phase~\cite{chen2020first,lavrentovich2020ferroelectric, sebastian2022ferroelectric}, where polar molecules with strong dipole moments exhibit both polar and nematic order, and the hexanematic phase, observed in Madin-Darby canine kidney cells, which displays both nematic and hexatic order~\cite{eckert2023hexanematic,armengol2023epithelia,armengol2024hydrodynamics}. A well-studied example of such coexistence is the tilted hexatic phase, which combines polar and hexatic order~\cite{bruinsma1982hexatic,dierker1987dynamics, sprunt1987light,selinger1989theory,selinger1991dynamics}. Similarly, a model of hexatic-nematic order coupling has been studied recently and found strings connecting three $+1/6$ hexatic topological defects, as well as an extended nematic $+1/2$ defect that connects the hexatic defects~\cite{drouin2022emergent}. These discoveries highlight the rich interplay between different types of orientational order in various physical systems, from synthetic liquid crystals to biological tissues, underscoring the importance of understanding these interactions for both fundamental science and practical applications.

Previous studies of coupled distinct order parameters have focused on the case of coupled scalar or vector order parameters~\cite{anisimov1981phase} and primarily on coupling the phases of orientational orders with fixed magnitudes~\cite{bruinsma1982hexatic,selinger1989theory,selinger1991dynamics,drouin2022emergent}, including gradient couplings as in flexoelectricity~\cite{meyer1969piezoelectric,shamid2013statistical,mertelj2018splay, sebastian2020ferroelectric, rosseto2020theory, sebastian2022ferroelectric, paik2024flexoelectricity}, as well as polar and nematic-like interactions of a single field~\cite{korshunov1985possible,lee1985strings,carpenter1989phase,shi2011boson,amiri2022unifying}. 

In this work, we propose a minimal theory that couples both the phase and magnitude of nematic and polar order parameters and captures interactions between distinct nematic and polar fields.  As we shall show, structural implications of accounting for this physical effect are drastic.\\

\noindent{\bf A minimal nematopolar model.} The three main contributions to the minimal nematopolar free energy that we consider are: (i) $\mathcal F_p$, an isotropic polar free energy, where the polar order parameter $\vec p$ has inherent tendency to be isotropic (ii) $\mathcal F_Q$, ordered nematic free energy, where the nematic order parameter $\overleftrightarrow{Q}$ (traceless symmetric rank-2 tensor) has inherent tendency to be ordered (iii) $\mathcal F_c$, a coupling which tends to align the polar and nematic order parameters.\footnote{A related polarnematic model, where the roles of $\vec p$ and $\overleftrightarrow{Q}$ are switched, i.e., $\vec p$ is ordered and $\overleftrightarrow{Q}$ isotropic, is analyzed in the Supplemental Material.} %Appendix~\ref{app:polarnematic}. } 
Then the total free energy $\mathcal F$ is the sum of these three
contributions, with $\mathcal F\left[\vec p,\overleftrightarrow{Q}\right] = \mathcal F_p\left[\vec p\right] + \mathcal F_Q\left[\overleftrightarrow{Q}\right] + \mathcal F_c\left[\vec p,\overleftrightarrow{Q}\right]$,
where
\begin{subequations}
	\begin{align}
		 \mathcal F_p &= \int d^2z\left[K_p |\nabla \vec p|^2 + \mathcal A_P |p|^2 \right] \label{eq:F_p} \\
		 \mathcal F_Q &= \int d^2z\left[K_Q \left|\nabla \cdot \overleftrightarrow{Q}\right|^2 + \mathcal A_Q \left(1-\left|Q\right|^2\right)^2 \right] \label{eq:F_Q}	 \\
		 \mathcal F_c &= \mathcal A_{NP} \int d^2z \left|\overleftrightarrow{Q} - \overleftrightarrow{P}\right|^2 , \label{eq:F_c}
	\end{align}
\end{subequations}
where explicitly for notational convenience, $|\nabla \vec p|^2 \equiv (\partial_i p^j)^2$, $|p|^2 \equiv p_i p^i$, $\left|\nabla \cdot \overleftrightarrow{Q}\right|^2 \equiv (\partial_i Q^{ij})^2$, $|Q|^2 \equiv \tr \left[\overleftrightarrow{Q}^2\right]$, $\overleftrightarrow{P} \equiv \vec p \vec p - (|p|^2/2)\overleftrightarrow{1}$, and $\left|\overleftrightarrow{Q} - \overleftrightarrow{P} \right|^2 \equiv \tr \left[\left(\overleftrightarrow{Q} - \overleftrightarrow{P}\right)^2\right]$.

In $\mathcal{F}_p$, the polar elasticity $K_p > 0$, with units of energy, defines the energetic cost of spatial variations of $\vec p$, while the disorder parameter $\mathcal{A}_P > 0$, with units of energy density, controls the strength of the polar order via the coherence length $w = \sqrt{K_p/\mathcal{A}_P}$, which governs spatial variations in the magnitude of $\vec p$. Similarly, in $\mathcal{F}_Q$, the nematic elasticity $K_Q > 0$, with units of energy, defines the energetic cost of spatial variations of $\overleftrightarrow{Q}$, and the order parameter $\mathcal{A}_Q > 0$, with units of energy density, controls the strength of the nematic order. By inspection of Eqs.~\eqref{eq:F_p} and \eqref{eq:F_Q}, $\vec p=0$ minimizes $\mathcal{F}_p$ and homogeneous $\overleftrightarrow{Q}$ with magnitude $\left|Q\right|=1$ minimizes $\mathcal{F}_Q$.
\begin{figure}[b!]
    \centering
    \includegraphics[width=\columnwidth]{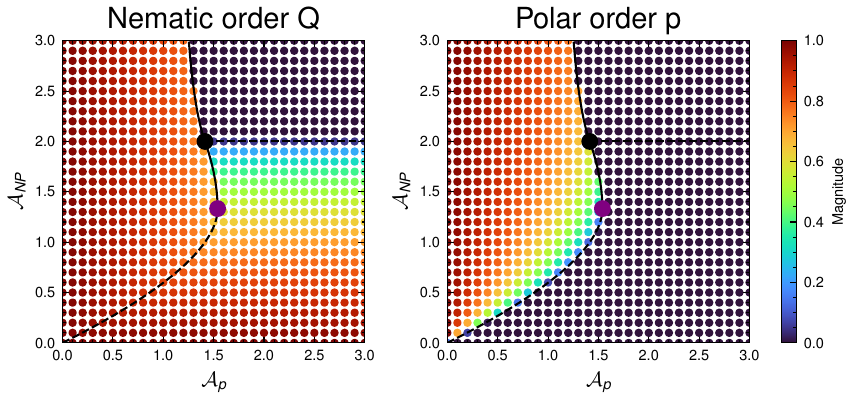}
    \caption{Phase diagram obtained from numerical simulations. Left panel: plot of $|Q|$. Right panel: plot of $|p|$. Each marker is a simulation result. Analytical phase boundaries and the critical points are superimposed on top, where the solid lines are discontinuous phase transitions, and the dashed lines are continuous phase transitions.}
    \label{fig:phaseD_sims}
\end{figure}

In $\mathcal F_c$, the aligning parameter $\mathcal A_{NP} > 0 $, with units of energy density, defines the energetic cost of misalignment between $\overleftrightarrow{Q}$ and $\overleftrightarrow{P}$. By inspection of Eq.~\eqref{eq:F_c}, $\overleftrightarrow{Q} = \overleftrightarrow{P}$ minimizes $\mathcal F_c$. The total free energy $\mathcal F$ is thus
\begin{align}
	\mathcal F[p,Q] &= \int d^2z\left[\mathcal A_{NP} \left|\overleftrightarrow{Q} - \overleftrightarrow{P}\right|^2 + \mathcal A_P |p|^2 + \mathcal A_Q (1-|Q|^2)^2 \right.\nonumber\\
	&\left.\qquad +K_p |\nabla \vec p|^2 + K_Q \left|\nabla \cdot \overleftrightarrow{Q}\right|^2 \right].
\end{align}
Without loss of generality, the polar vector $\vec p$ is assumed to align along the $x$-direction with magnitude $p$, while the nematic order parameter $\overleftrightarrow{Q}$ is aligned with the polar order with magnitude $Q$, and both are homogeneous. The free energy is then expressed as 
\beq 
\mathcal{F} = \mathcal{A}_{NP} (Q - p^2)^2 + \mathcal{A}_P p^2 + \mathcal{A}_Q (1-Q^2)^2, 
\eeq
where $\mathcal{A}_P$, $\mathcal{A}_Q$, and $\mathcal{A}_{NP}$ represent contributions from polar disorder, nematic order, and polar-nematic coupling, respectively. By rescaling $\mathcal{F}$, we set $\mathcal{A}_Q = 1$ without loss of generality. 

The competition between the three terms leads to three distinct phases, as shown in the phase diagram (Fig.~\ref{fig:phase_diagram}). The $\mathcal{A}_P$ term suppresses polar order ($p=0$), the $\mathcal{A}_Q$ term favors maximal nematic order ($Q=1$), and the $\mathcal{A}_{NP}$ term couples $Q$ and $p$ via $Q = p^2$. Two limiting cases provide insight into the phase behavior: (i) $\mathcal{A}_{NP} \gg 1$ and (ii) $\mathcal{A}_P \gg 1$. 

(i) For $\mathcal{A}_{NP} \gg 1$, the coupling enforces $Q \sim p^2$. When $\mathcal{A}_P$ is large, $p=0$ and $Q=0$, corresponding to the isotropic phase (red region, Fig.~\ref{fig:phase_diagram}). For small $\mathcal{A}_P$ ($\mathcal{A}_P < 4\sqrt{6}/9$), nematic ordering dominates, leading to $Q \neq 0$ and $Q \sim p^2 \neq 0$, defining the nematopolar phase (cyan region). These phases are separated by a first-order phase transition (solid curve), asymptoting to $\mathcal{A}_P = 4\sqrt{6}/9$.

(ii) For $\mathcal{A}_P \gg 1$, polar disorder dominates, yielding $p=0$. If $\mathcal{A}_{NP}$ exceeds 2, the system remains isotropic ($Q = p^2 = 0$). For $\mathcal{A}_{NP} < 2$, nematic order emerges ($Q \neq 0$, $p=0$), defining the nematic phase (green region). The isotropic and nematic phases are separated by a continuous transition (dashed line, Fig.~\ref{fig:phase_diagram}), which satisfies $\mathcal{A}_{NP} = 2$ for $\mathcal{A}_P \geq \sqrt{2}$. 

The nematopolar and nematic phases are delineated by a boundary comprising a continuous transition (dashed curve, origin to purple tricritical point) and a first-order transition (solid curve, tricritical to black triple point). The dashed boundary satisfies 
\beq 
2\mathcal{A}_{NP}\sqrt{1 - \mathcal{A}_{NP}/2} = \mathcal{A}_P \label{eq:continuous}, 
\eeq 
with the critical $\mathcal{A}_{NP}^*(\mathcal{A}_P)$ defined as its smallest positive solution. The analytical results are corroborated by numerical simulations in Fig.~\ref{fig:phaseD_sims}, with details in the Supplemental Material.\\%Appendix~\ref{app:numerics}.
\begin{table}[]
\begin{tabular}{l |l | l | l}
         & $\alpha$ & $\beta$  & $\gamma$  \\
         \hline
N $\to$ I: $Q$      & 0        & 1/2      & 1         \\
         \hline

NP $\to$ N: $p$      & (0, 1)   & (1/2, 1/4) & (1, 1/2)      \\
         \hline
N $\to$ NP: $\sigma$ & 0        & 1        & 0       
\end{tabular}
\caption{Mean-field critical exponents. The critical exponents $\alpha$, $\beta$, and $\gamma$ are defined by specific heat $C \equiv -T \p{^2\mathcal F}{T^2}\propto (T - T^*)^{-\alpha}$, magnetization $m \propto (T - T^*)^{\beta}$, and susceptibility $\chi \equiv \left.\p{m}{h}\right|_{h=0} \propto (T - T^*)^{-\gamma}$, respectively, where $h$ is an external field. The second entries for $p$ denote the shifted values at the tricritical point.}
\label{tab:crit_exponents}
\end{table}

\noindent{\bf Critical exponents and a new universality class.} The phases and phase transitions identified are characterized by analytically obtained mean-field critical exponents for the N-I and N-NP continuous phase transitions, summarized in Table~\ref{tab:crit_exponents} (see the Supplemental Material for derivations).  %(see Appendix~\ref{app:crit_exps} for derivations). 
The critical exponents $\alpha$, $\beta$, and $\gamma$ describe the behavior of the specific heat $C \equiv -T \partial^2\mathcal{F}/\partial T^2 \propto (T - T^*)^{-\alpha}$, the order parameter $m \propto (T - T^*)^{\beta}$, and the susceptibility $\chi \equiv \left.\partial m / \partial h\right|_{h=0} \propto (T - T^*)^{-\gamma}$, where $h$ is an external field. 

For the N-I phase boundary, the scaling $2 - \mathcal{A}_{NP} \approx T^* - T$ is assumed, with $m = Q$. This yields critical exponents identical to the mean-field Ising model: $\alpha = 0$, $\beta = 1/2$, and $\gamma = 1$. For the N-NP phase boundary, the scaling $\mathcal{A}_P^* - \mathcal{A}_P \approx T - T^*$ is used, with two relevant order parameters: $p$ and the locking order parameter $\sigma \equiv Q - p^2 - \mathcal{A}_P/(2\mathcal{A}_{NP})$, where $\sigma = 0$ in the nematopolar phase. 

Generically, $p$ exhibits critical exponents $\alpha = 0$, $\beta = 1/2$, and $\gamma = 1$. At the tricritical point $(\mathcal{A}_P, \mathcal{A}_{NP}) = (8/\sqrt{27}, 4/3)$, these exponents shift to $\alpha = 1$, $\beta = 1/4$, and $\gamma = 1/2$. For $\sigma$, the exponents are $\alpha = 0$, $\beta = 1$, and $\gamma = 0$. Notably, these exponents for the N-NP boundary do not correspond to standard universality classes of nematic or polar matter. 

The change in $\alpha$ at the tricritical point reflects the discontinuity of $\partial^2\mathcal{F}/\partial T^2$ across the transition. This finding highlights the unconventional nature of critical phenomena in the system and underscores the richness of the underlying phase structure.\\

\begin{figure*}[t]
	\centering
	\subfloat[Sketch]{\includegraphics[width=.32\linewidth]{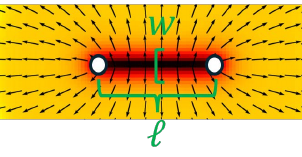}}
	\hfill
	\subfloat[Confined: $\mathcal A_{NP}> \mathcal A_{NP}^*$]{\includegraphics[width=.32\linewidth]{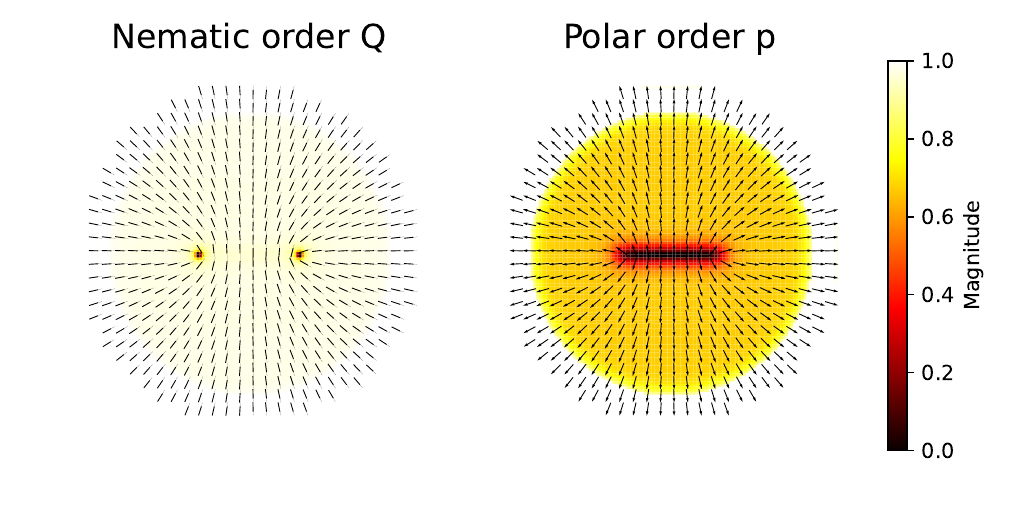}}
	\label{fig:confined}
	\hfill
	\subfloat[Deconfined: $\mathcal A_{NP} < \mathcal A_{NP}^*$]{\includegraphics[width=.32\linewidth]{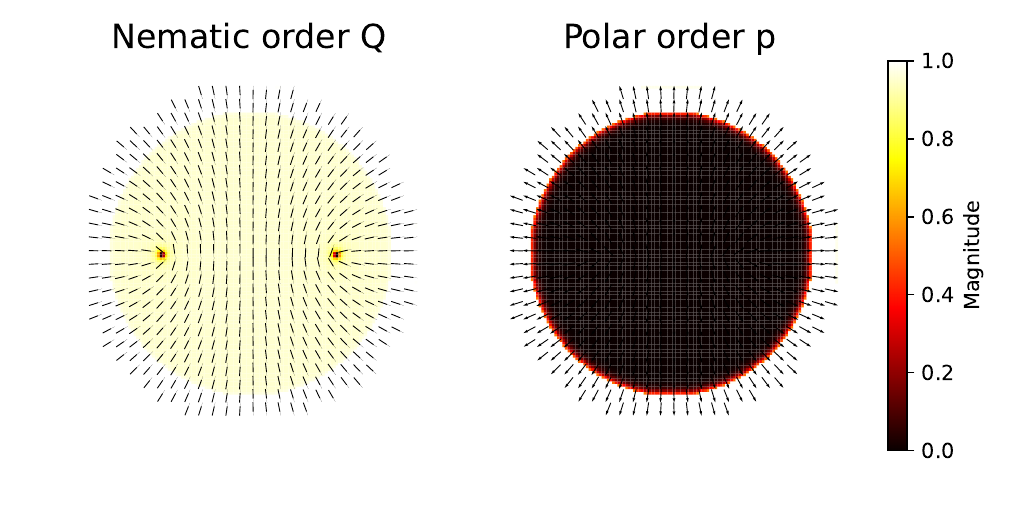}}
	\label{fig:deconfined} \\
	\subfloat[]{\includegraphics[width=.32\linewidth]{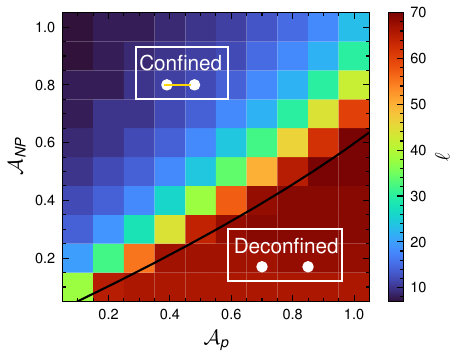}}
	%\caption{Phase diagram with data from numerical simulations of confinement/deconfinement transition. Color indicates separation distance $\ell$ of the two $+1/2$ defects.}
	\label{fig:l}
	\hfill
	\subfloat[]{\includegraphics[width=.32\linewidth]{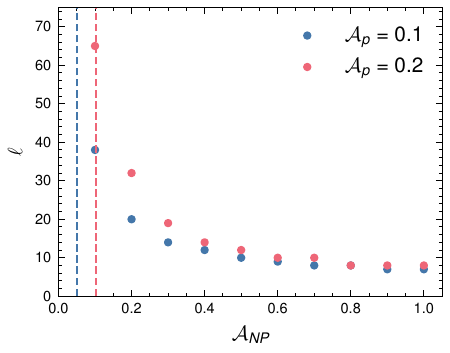}}
	%\label{fig:}
	\hfill
	\subfloat[]{\includegraphics[width=.32\linewidth]{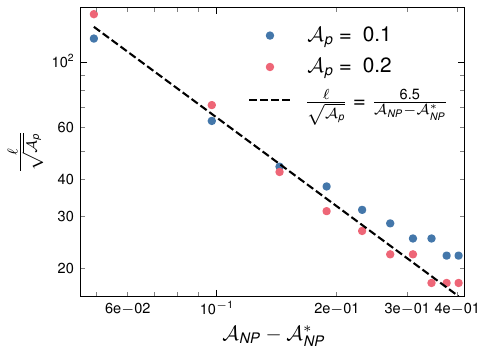}}
	%\label{fig:}
	%\caption{Plots of equilibrium size of bound pairs of $+1/2$ defects for small values of $\mathcal A_P$. (a) Colored dashed vertical lines denote $\mathcal A_{NP}^*$. (b) Log-log plot, where red dashed line is fit of Eq.XXX.}
	%\caption{Stationary plots of $p$ and $Q$ where color denotes the magnitude for (a) confined regime and (b) deconfined regime.}
	\caption{Confining strings. (a) Sketch of elongated $+1$ polar defect of length $\ell$ and width $w$. A pair of $+1/2$ defects, at the endpoints of the string, are confined. (b) and (c) Stationary plots of $p$ and $Q$ where color denotes the magnitude for (b) confined regime ($\mathcal A_P = 0.1$ and $\mathcal A_{NP} = 0.1$) and (c) deconfined regime ($\mathcal A_P = 1.0$ and $\mathcal A_{NP} = 0.1$). (d) Phase diagram with data from numerical simulations of confinement/deconfinement transition, where color indicates separation distance $\ell$ of the two $+1/2$ defects. Insets indicate parameter region of confined and deconfined defects. (e) and (f) Plots of equilibrium length $\ell$ of bound pairs of $+1/2$ defects for small values of $\mathcal A_P$. (e) Colored dashed vertical lines denote $\mathcal A_{NP}^*$. (f) Log-log plot, where red dashed line is fit of Eq.~\eqref{eq:ell}.}
	\label{fig:string}
\end{figure*}

%\subsection{Size of confining string}
\noindent{\bf Confining strings.} Having established the distinct ordered phases and their transitions, we analyze the behavior of topological defects, where order breaks down. For analytical clarity, we focus on isolated defects. Consider a configuration with a $+1$ polar defect and a pair of $+1/2$ nematic defects (not necessarily bound). In nematics, $+1$ defects decompose into pairs of $+1/2$ defects due to Coulomb repulsion. For polar order, however, the elementary defect charge is $+1$. If nematic order $\overleftrightarrow{Q}$ is strongly coupled to polar order $\vec p$, such that $\overleftrightarrow{Q} \approx \overleftrightarrow{P}$, a $+1$ polar defect induces a minimal $+1$ nematic defect, leading to confinement of $+1/2$ nematic defects. The strength of this confinement depends on the coupling parameter $\mathcal{A}_{NP}$.

In the strongly locked regime ($\mathcal{A}_{NP} \gg \mathcal{A}_{NP}^*$), $\overleftrightarrow{Q} \approx \overleftrightarrow{P}$, confining $+1/2$ nematic defects. In the weakly locked regime ($\mathcal{A}_{NP} \gtrsim \mathcal{A}_{NP}^*$), frustration emerges between nematic and polar fields. This frustration localizes along a string connecting the $+1/2$ defects, where $\vec p$ becomes ordered, resulting in a stretched defect core. The string has a characteristic width $w$, corresponding to the polar coherence length, and a length $\ell$, determined by minimizing the free energy.

The free energy components include the nematic elastic term 
\beq \mathcal{F}_Q = \frac{\pi}{2} K_Q \ln (L/\ell), \eeq
where $L$ is the system size, and the combined contributions of polar and coupling energies, given by $\mathcal{F}_c + \mathcal{F}_p \sim f(\mathcal{A}_{NP}) w \ell$. For $\mathcal{A}_{NP} \gg 1$, $f(\mathcal{A}_{NP}) \sim \mathcal{A}_{NP}$. Near the N-NP curve, the string tension must vanish, requiring $f(\mathcal{A}_{NP}) \sim \mathcal{A}_{NP} - \mathcal{A}_{NP}^*(\mathcal{A}_P)$, where $\mathcal{A}_{NP}^*(\mathcal{A}_P)$ lies on the N-NP curve. Using $w = \sqrt{K_p/\mathcal{A}_P}$, the string tension is 
\beq 
T \equiv (\mathcal{A}_{NP} - \mathcal{A}_{NP}^*) \sqrt{\frac{K_p}{\mathcal{A}_P}}. 
\eeq
Replacing this into the total free energy 
\beq 
\mathcal{F} \sim \frac{\pi}{2} K_Q \ln (L/\ell) + T \ell, 
\eeq
and minimizing with respect to the string length $\ell$ gives 
\beq 
\ell \sim \frac{\pi}{2} \sqrt{\frac{K_Q^2 \mathcal{A}_P}{K_p}} \frac{1}{\mathcal{A}_{NP} - \mathcal{A}_{NP}^*} \label{eq:ell}. 
\eeq

Numerical simulations confirm these predictions, as shown in Fig.~\ref{fig:string}. Panels (b) and (c) illustrate steady-state configurations in the confined and deconfined regimes, respectively. Panel (d) measures the equilibrium string length $\ell$, and Fig.~\ref{fig:string}(e-f) verifies the theoretical scaling for small $\mathcal{A}_P$. These findings reveal how defect confinement arises from the interplay between polar and nematic orders, with implications for understanding topological excitations in systems with coexisting orientational orders.

Unlike recent experimental observations of strings connecting neutral nematic defect pairs in endothelial cell layers~\cite{ruider2024topological}, our model reveals the confinement of same-sign nematic defects. A neutral nematic defect pair in the nematopolar phase would similarly form a string, but this configuration is unstable due to the combined effect of Coulombic and string tension forces, which are both attractive. In contrast, for same-sign defects, the Coulombic repulsion opposes the string tension, resulting in a stable equilibrium. This key distinction offers fresh insights into the stabilization mechanisms of topological defects and raises intriguing questions about the potential role of external fields or other perturbations in stabilizing strings for neutral defect pairs.\\

\noindent{\bf Coupling to External Fields.} The impact of external fields on the system is examined by coupling $\overleftrightarrow{Q}$ and $\vec p$ to fields $\overleftrightarrow{H}$ and $\vec h$, respectively, through the following free energy:  
\beq  
\mathcal F = \mathcal A_{NP} \left|\overleftrightarrow{Q} - \overleftrightarrow{P} \right|^2 + \mathcal A_P|p|^2 + (|Q|^2-1)^2 -\tr [\overleftrightarrow{H}\overleftrightarrow{Q}] - \vec h\cdot \vec p . \label{eq:F_ext}  
\eeq  
Two cases are considered: (i) $\vec h = 0, \overleftrightarrow{H} \neq 0$, and (ii) $\vec h \neq 0, \overleftrightarrow{H} = 0$  (see the Supplemental Material for derivations).\\ %(see Appendix~\ref{app:ext_fields} for details).\\
(i) When $\vec h = 0$, the free energy is minimized when $\overleftrightarrow{H}$, $\overleftrightarrow{Q}$, and $\overleftrightarrow{P}$ are all aligned. Non-zero external field $\overleftrightarrow{H}$ prohibits $\overleftrightarrow{Q} = 0$ as a solution, eliminating the nematic-isotropic (N-I) phase boundary. However, both $\vec p = 0$ and $\vec p \neq 0$ remain valid solutions, preserving the continuity of the nematic-nematopolar (N-NP) phase boundary, albeit in a deformed form.\\
(ii) When $\overleftrightarrow{H} = 0$, the free energy is minimized when $\vec p$ aligns with $\vec h$ and $\overleftrightarrow{Q}$ aligns with $\overleftrightarrow{P}$. Here, non-zero $\vec h$ prohibits $\vec p = 0$ as a solution, thereby also excluding $\overleftrightarrow{Q} = 0$. As a result, the nematic and isotropic phases are no longer stable, and only the nematopolar phase persists.

The fate of topological defects in the presence of a large external field $\vec h$ is particularly intriguing. Consider a region where $\vec h$ points in the $+\hat{x}$ direction. For a $+1$ polar aster defect at the origin, the field $p^x$ changes sign across $x=0$, leading to an asymmetric energy contribution. In the region $x > 0$, $\vec p$ aligns with $\vec h$ and thus lowers the energy, whereas in the region $x < 0$, $\vec p$ misaligns and thus raises the energy. This energy asymmetry induces a force on the defect, driving it in the $-x$ direction to align with $\vec h$. Thus, external fields render defects mobile and, for sufficiently large $\vec h$, repel them entirely  (see the Supplemental Material for representative snapshots of movement of the topological defects). %(see Fig.~\ref{fig:expel} in Appendix~\ref{app:numerics} for representative snapshots of movement of the topological defects). 
This result highlights the profound influence of external fields in suppressing topological defects, fundamentally altering the phase behavior and providing new avenues to control systems with coexisting orientational orders.\\

\begin{figure}[t]
	\centering
	\subfloat[$\overleftrightarrow{H} \neq 0$]{\includegraphics[width=.49\columnwidth]{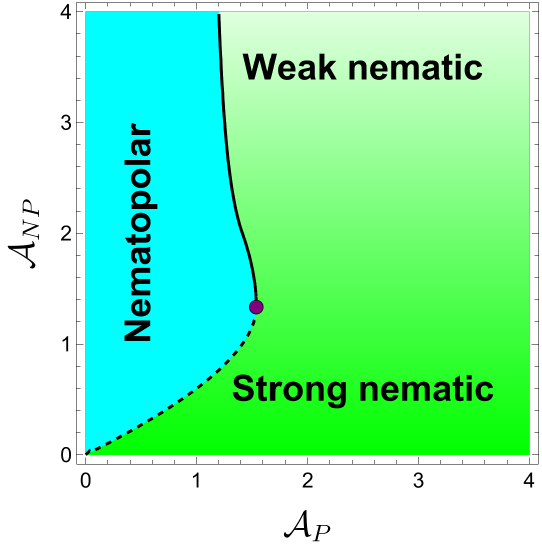}}
	\hfill
	%\caption{Schematic comparison of undeformed ($H = 0$) and deformed ($H \neq 0$) phase diagrams for $H \ll 1$.}
	\subfloat[$\vec h \neq 0$]{\includegraphics[width=.49\columnwidth]{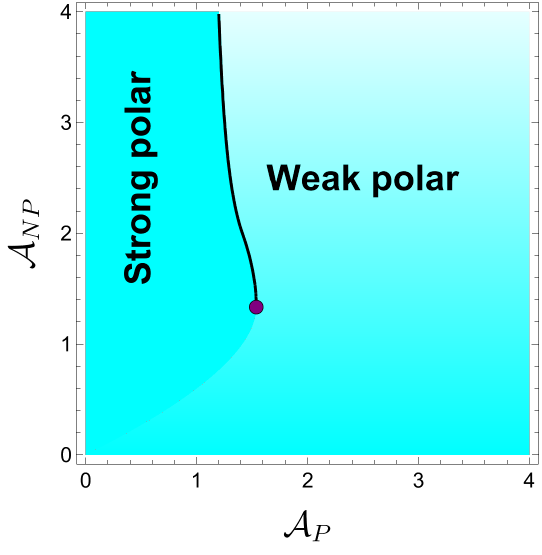}}
	\caption{Schematic phase diagrams in the presence of external fields. (a) $\overleftrightarrow{Q}$ is coupled to small external field $\overleftrightarrow{H}\neq 0$. The N-I phase boundary disappears, but the other phase boundaries remain, but deformed. (b)  $\vec p$ is coupled to small external field $\vec h \neq 0$. Both continuous phase boundaries disappear, leaving only a single ordered phase.}
\end{figure}	
\noindent{\bf Discussion.}  
Introducing a minimal nematopolar model that captures the coupling between polar and nematic order, and through analytical calculations, we uncovered a rich phase diagram featuring nematopolar, nematic, and isotropic phases, with phase transitions characterized by distinct critical exponents that do not fall under any known universality classes. These theoretical findings were validated by extensive numerical simulations.  

The nematopolar phase presents dual and complementary perspectives: (i) nematic order induces polar order, and (ii) a polar $+1$ defect confines a pair of nematic $+1/2$ defects. In this confined phase, the $+1/2$ defects are connected by a string with a well-defined tension, proportional to $\mathcal A_{NP} - \mathcal A_{NP}^*$. This string stabilizes the defect pair, and its equilibrium length $\ell$ scales as $\ell \sim 1/(\mathcal A_{NP} - \mathcal A_{NP}^*)$. Such confinement and string formation underscore the intricate coupling between polar and nematic fields and provide a novel mechanism for defect stabilization in active and passive systems.  
%
%Unlike recent experimental observations of strings connecting neutral nematic defect pairs in endothelial cell layers~\cite{ruider2024topological}, our model reveals the confinement of same-sign nematic defects. A neutral nematic defect pair in the nematopolar phase would similarly form a string, but this configuration is unstable due to the combined effect of Coulombic and string tension forces, which are both attractive. In contrast, for same-sign defects, the Coulombic repulsion opposes the string tension, resulting in a stable equilibrium. This key distinction offers fresh insights into the stabilization mechanisms of topological defects and raises intriguing questions about the potential role of external fields or other perturbations in stabilizing strings for neutral defect pairs.  
%In this vein, our results reveal the profound impact of external fields on phase behavior and defect dynamics. Fields coupled to the polar or nematic order destabilize isotropic and nematic phases, leaving only the nematopolar phase. These fields can repel topological defects, suppressing their formation and rendering systems free of defects in strong-field regions. This ability to control defect dynamics could have far-reaching implications for materials design and active matter systems.

Our findings highlight the rich phenomenology that emerges from the interplay between nematic and polar orientational order, offering new insights into systems with competing symmetries. The identification of distinct phases and critical behaviors, including novel universality at the nematopolar-nematic transition, broadens the landscape of phase transitions in soft and active matter. The identification of string-mediated confinement of nematic defects in the nematopolar phase provides a striking example of frustration-driven defect interactions, with potential implications for the design and control of defect networks in liquid crystals and active materials. Collectively, these results pave the way for understanding how coupled orientational fields shape collective behavior in diverse soft matter systems, with potential applications ranging from liquid crystal technologies to biological systems with intrinsic order.

\section*{Acknowledgments}
It is a pleasure to acknowledge helpful conversations with Aboutaleb Amiri, Jonathan Bauermann, Lara Braverman, Nikta Fakhri, Luca Giomi, Mehran Kardar, and David Nelson. This work is partially supported by the Center for Mathematical Sciences and Applications at Harvard University (F. V.). F.V. gratefully acknowledges the Niels Bohr Institute for their hospitality, where some of the research for this paper was performed. A. D. acknowledges funding from the Novo Nordisk Foundation (grant No. NNF18SA0035142 and NERD grant No. NNF21OC0068687), Villum Fonden (Grant no. 29476), and the European Union (ERC, PhysCoMeT, 101041418). Views and opinions expressed are however those of the authors only and do not necessarily reflect those of the European Union or the European Research Council. Neither the European Union nor the granting authority can be held responsible for them.
	
	%\section*{Appendix}
	
	%\appendix
	
	\bibliography{refs}

\begin{thebibliography}{38}%
\makeatletter
\providecommand \@ifxundefined [1]{%
 \@ifx{#1\undefined}
}%
\providecommand \@ifnum [1]{%
 \ifnum #1\expandafter \@firstoftwo
 \else \expandafter \@secondoftwo
 \fi
}%
\providecommand \@ifx [1]{%
 \ifx #1\expandafter \@firstoftwo
 \else \expandafter \@secondoftwo
 \fi
}%
\providecommand \natexlab [1]{#1}%
\providecommand \enquote  [1]{``#1''}%
\providecommand \bibnamefont  [1]{#1}%
\providecommand \bibfnamefont [1]{#1}%
\providecommand \citenamefont [1]{#1}%
\providecommand \href@noop [0]{\@secondoftwo}%
\providecommand \href [0]{\begingroup \@sanitize@url \@href}%
\providecommand \@href[1]{\@@startlink{#1}\@@href}%
\providecommand \@@href[1]{\endgroup#1\@@endlink}%
\providecommand \@sanitize@url [0]{\catcode `\\12\catcode `\$12\catcode `\&12\catcode `\#12\catcode `\^12\catcode `\_12\catcode `\%12\relax}%
\providecommand \@@startlink[1]{}%
\providecommand \@@endlink[0]{}%
\providecommand \url  [0]{\begingroup\@sanitize@url \@url }%
\providecommand \@url [1]{\endgroup\@href {#1}{\urlprefix }}%
\providecommand \urlprefix  [0]{URL }%
\providecommand \Eprint [0]{\href }%
\providecommand \doibase [0]{https://doi.org/}%
\providecommand \selectlanguage [0]{\@gobble}%
\providecommand \bibinfo  [0]{\@secondoftwo}%
\providecommand \bibfield  [0]{\@secondoftwo}%
\providecommand \translation [1]{[#1]}%
\providecommand \BibitemOpen [0]{}%
\providecommand \bibitemStop [0]{}%
\providecommand \bibitemNoStop [0]{.\EOS\space}%
\providecommand \EOS [0]{\spacefactor3000\relax}%
\providecommand \BibitemShut  [1]{\csname bibitem#1\endcsname}%
\let\auto@bib@innerbib\@empty
%</preamble>
\bibitem [{\citenamefont {de~Gennes}\ and\ \citenamefont {Prost}(1993)}]{gennes1993the}%
  \BibitemOpen
  \bibfield  {author} {\bibinfo {author} {\bibfnamefont {P.~G.}\ \bibnamefont {de~Gennes}}\ and\ \bibinfo {author} {\bibfnamefont {J.}~\bibnamefont {Prost}},\ }\href@noop {} {\emph {\bibinfo {title} {The physics of liquid crystals}}}\ (\bibinfo  {publisher} {Clarendon Press, Oxford},\ \bibinfo {year} {1993})\BibitemShut {NoStop}%
\bibitem [{\citenamefont {Yang}\ and\ \citenamefont {Wu}(2014)}]{yang2014fundamentals}%
  \BibitemOpen
  \bibfield  {author} {\bibinfo {author} {\bibfnamefont {D.-K.}\ \bibnamefont {Yang}}\ and\ \bibinfo {author} {\bibfnamefont {S.-T.}\ \bibnamefont {Wu}},\ }\href@noop {} {\emph {\bibinfo {title} {Fundamentals of liquid crystal devices}}}\ (\bibinfo  {publisher} {John Wiley \& Sons},\ \bibinfo {year} {2014})\BibitemShut {NoStop}%
\bibitem [{\citenamefont {Doostmohammadi}\ and\ \citenamefont {Ladoux}(2021)}]{doostmohammadi2021physics}%
  \BibitemOpen
  \bibfield  {author} {\bibinfo {author} {\bibfnamefont {A.}~\bibnamefont {Doostmohammadi}}\ and\ \bibinfo {author} {\bibfnamefont {B.}~\bibnamefont {Ladoux}},\ }\bibfield  {title} {\bibinfo {title} {Physics of liquid crystals in cell biology},\ }\href@noop {} {\bibfield  {journal} {\bibinfo  {journal} {Trends in cell biology}\ } (\bibinfo {year} {2021})}\BibitemShut {NoStop}%
\bibitem [{\citenamefont {Marchetti}\ \emph {et~al.}(2013)\citenamefont {Marchetti}, \citenamefont {Joanny}, \citenamefont {Ramaswamy}, \citenamefont {Liverpool}, \citenamefont {Prost}, \citenamefont {Rao},\ and\ \citenamefont {Simha}}]{marchetti2013hydrodynamics}%
  \BibitemOpen
  \bibfield  {author} {\bibinfo {author} {\bibfnamefont {M.~C.}\ \bibnamefont {Marchetti}}, \bibinfo {author} {\bibfnamefont {J.-F.}\ \bibnamefont {Joanny}}, \bibinfo {author} {\bibfnamefont {S.}~\bibnamefont {Ramaswamy}}, \bibinfo {author} {\bibfnamefont {T.~B.}\ \bibnamefont {Liverpool}}, \bibinfo {author} {\bibfnamefont {J.}~\bibnamefont {Prost}}, \bibinfo {author} {\bibfnamefont {M.}~\bibnamefont {Rao}},\ and\ \bibinfo {author} {\bibfnamefont {R.~A.}\ \bibnamefont {Simha}},\ }\bibfield  {title} {\bibinfo {title} {Hydrodynamics of soft active matter},\ }\href@noop {} {\bibfield  {journal} {\bibinfo  {journal} {Reviews of modern physics}\ }\textbf {\bibinfo {volume} {85}},\ \bibinfo {pages} {1143} (\bibinfo {year} {2013})}\BibitemShut {NoStop}%
\bibitem [{\citenamefont {Kaganer}\ \emph {et~al.}(1999)\citenamefont {Kaganer}, \citenamefont {M{\"o}hwald},\ and\ \citenamefont {Dutta}}]{kaganer1999structure}%
  \BibitemOpen
  \bibfield  {author} {\bibinfo {author} {\bibfnamefont {V.~M.}\ \bibnamefont {Kaganer}}, \bibinfo {author} {\bibfnamefont {H.}~\bibnamefont {M{\"o}hwald}},\ and\ \bibinfo {author} {\bibfnamefont {P.}~\bibnamefont {Dutta}},\ }\bibfield  {title} {\bibinfo {title} {Structure and phase transitions in langmuir monolayers},\ }\href@noop {} {\bibfield  {journal} {\bibinfo  {journal} {Reviews of Modern Physics}\ }\textbf {\bibinfo {volume} {71}},\ \bibinfo {pages} {779} (\bibinfo {year} {1999})}\BibitemShut {NoStop}%
\bibitem [{\citenamefont {Schenning}\ \emph {et~al.}(2017)\citenamefont {Schenning}, \citenamefont {Crawford},\ and\ \citenamefont {Broer}}]{schenning2017liquid}%
  \BibitemOpen
  \bibfield  {author} {\bibinfo {author} {\bibfnamefont {A.}~\bibnamefont {Schenning}}, \bibinfo {author} {\bibfnamefont {G.~P.}\ \bibnamefont {Crawford}},\ and\ \bibinfo {author} {\bibfnamefont {D.~J.}\ \bibnamefont {Broer}},\ }\href@noop {} {\emph {\bibinfo {title} {Liquid crystal sensors}}}\ (\bibinfo  {publisher} {CRC Press},\ \bibinfo {year} {2017})\BibitemShut {NoStop}%
\bibitem [{\citenamefont {Ula}\ \emph {et~al.}(2018)\citenamefont {Ula}, \citenamefont {Traugutt}, \citenamefont {Volpe}, \citenamefont {Patel}, \citenamefont {Yu},\ and\ \citenamefont {Yakacki}}]{ula2018liquid}%
  \BibitemOpen
  \bibfield  {author} {\bibinfo {author} {\bibfnamefont {S.~W.}\ \bibnamefont {Ula}}, \bibinfo {author} {\bibfnamefont {N.~A.}\ \bibnamefont {Traugutt}}, \bibinfo {author} {\bibfnamefont {R.~H.}\ \bibnamefont {Volpe}}, \bibinfo {author} {\bibfnamefont {R.~R.}\ \bibnamefont {Patel}}, \bibinfo {author} {\bibfnamefont {K.}~\bibnamefont {Yu}},\ and\ \bibinfo {author} {\bibfnamefont {C.~M.}\ \bibnamefont {Yakacki}},\ }\bibfield  {title} {\bibinfo {title} {Liquid crystal elastomers: an introduction and review of emerging technologies},\ }\href@noop {} {\bibfield  {journal} {\bibinfo  {journal} {Liquid Crystals Reviews}\ }\textbf {\bibinfo {volume} {6}},\ \bibinfo {pages} {78} (\bibinfo {year} {2018})}\BibitemShut {NoStop}%
\bibitem [{\citenamefont {Volfson}\ \emph {et~al.}(2008)\citenamefont {Volfson}, \citenamefont {Cookson}, \citenamefont {Hasty},\ and\ \citenamefont {Tsimring}}]{volfson2008biomechanical}%
  \BibitemOpen
  \bibfield  {author} {\bibinfo {author} {\bibfnamefont {D.}~\bibnamefont {Volfson}}, \bibinfo {author} {\bibfnamefont {S.}~\bibnamefont {Cookson}}, \bibinfo {author} {\bibfnamefont {J.}~\bibnamefont {Hasty}},\ and\ \bibinfo {author} {\bibfnamefont {L.~S.}\ \bibnamefont {Tsimring}},\ }\bibfield  {title} {\bibinfo {title} {Biomechanical ordering of dense cell populations},\ }\href@noop {} {\bibfield  {journal} {\bibinfo  {journal} {Proceedings of the National Academy of Sciences}\ }\textbf {\bibinfo {volume} {105}},\ \bibinfo {pages} {15346} (\bibinfo {year} {2008})}\BibitemShut {NoStop}%
\bibitem [{\citenamefont {Doostmohammadi}\ \emph {et~al.}(2016)\citenamefont {Doostmohammadi}, \citenamefont {Thampi},\ and\ \citenamefont {Yeomans}}]{doostmohammadi2016defect}%
  \BibitemOpen
  \bibfield  {author} {\bibinfo {author} {\bibfnamefont {A.}~\bibnamefont {Doostmohammadi}}, \bibinfo {author} {\bibfnamefont {S.~P.}\ \bibnamefont {Thampi}},\ and\ \bibinfo {author} {\bibfnamefont {J.~M.}\ \bibnamefont {Yeomans}},\ }\bibfield  {title} {\bibinfo {title} {Defect-mediated morphologies in growing cell colonies},\ }\href@noop {} {\bibfield  {journal} {\bibinfo  {journal} {Physical review letters}\ }\textbf {\bibinfo {volume} {117}},\ \bibinfo {pages} {048102} (\bibinfo {year} {2016})}\BibitemShut {NoStop}%
\bibitem [{\citenamefont {Kruse}\ \emph {et~al.}(2005)\citenamefont {Kruse}, \citenamefont {Joanny}, \citenamefont {J{\"u}licher}, \citenamefont {Prost},\ and\ \citenamefont {Sekimoto}}]{kruse2005generic}%
  \BibitemOpen
  \bibfield  {author} {\bibinfo {author} {\bibfnamefont {K.}~\bibnamefont {Kruse}}, \bibinfo {author} {\bibfnamefont {J.-F.}\ \bibnamefont {Joanny}}, \bibinfo {author} {\bibfnamefont {F.}~\bibnamefont {J{\"u}licher}}, \bibinfo {author} {\bibfnamefont {J.}~\bibnamefont {Prost}},\ and\ \bibinfo {author} {\bibfnamefont {K.}~\bibnamefont {Sekimoto}},\ }\bibfield  {title} {\bibinfo {title} {Generic theory of active polar gels: a paradigm for cytoskeletal dynamics},\ }\href@noop {} {\bibfield  {journal} {\bibinfo  {journal} {The European Physical Journal E}\ }\textbf {\bibinfo {volume} {16}},\ \bibinfo {pages} {5} (\bibinfo {year} {2005})}\BibitemShut {NoStop}%
\bibitem [{\citenamefont {Saw}\ \emph {et~al.}(2017)\citenamefont {Saw}, \citenamefont {Doostmohammadi}, \citenamefont {Nier}, \citenamefont {Kocgozlu}, \citenamefont {Thampi}, \citenamefont {Toyama}, \citenamefont {Marcq}, \citenamefont {Lim}, \citenamefont {Yeomans},\ and\ \citenamefont {Ladoux}}]{saw2017topological}%
  \BibitemOpen
  \bibfield  {author} {\bibinfo {author} {\bibfnamefont {T.~B.}\ \bibnamefont {Saw}}, \bibinfo {author} {\bibfnamefont {A.}~\bibnamefont {Doostmohammadi}}, \bibinfo {author} {\bibfnamefont {V.}~\bibnamefont {Nier}}, \bibinfo {author} {\bibfnamefont {L.}~\bibnamefont {Kocgozlu}}, \bibinfo {author} {\bibfnamefont {S.}~\bibnamefont {Thampi}}, \bibinfo {author} {\bibfnamefont {Y.}~\bibnamefont {Toyama}}, \bibinfo {author} {\bibfnamefont {P.}~\bibnamefont {Marcq}}, \bibinfo {author} {\bibfnamefont {C.~T.}\ \bibnamefont {Lim}}, \bibinfo {author} {\bibfnamefont {J.~M.}\ \bibnamefont {Yeomans}},\ and\ \bibinfo {author} {\bibfnamefont {B.}~\bibnamefont {Ladoux}},\ }\bibfield  {title} {\bibinfo {title} {Topological defects in epithelia govern cell death and extrusion},\ }\href {https://doi.org/10.1038/nature21718} {\bibfield  {journal} {\bibinfo  {journal} {Nature}\ }\textbf {\bibinfo {volume} {544}},\ \bibinfo {pages} {212–216} (\bibinfo {year} {2017})}\BibitemShut {NoStop}%
\bibitem [{\citenamefont {Chen}\ \emph {et~al.}(2020)\citenamefont {Chen}, \citenamefont {Korblova}, \citenamefont {Dong}, \citenamefont {Wei}, \citenamefont {Shao}, \citenamefont {Radzihovsky}, \citenamefont {Glaser}, \citenamefont {Maclennan}, \citenamefont {Bedrov}, \citenamefont {Walba} \emph {et~al.}}]{chen2020first}%
  \BibitemOpen
  \bibfield  {author} {\bibinfo {author} {\bibfnamefont {X.}~\bibnamefont {Chen}}, \bibinfo {author} {\bibfnamefont {E.}~\bibnamefont {Korblova}}, \bibinfo {author} {\bibfnamefont {D.}~\bibnamefont {Dong}}, \bibinfo {author} {\bibfnamefont {X.}~\bibnamefont {Wei}}, \bibinfo {author} {\bibfnamefont {R.}~\bibnamefont {Shao}}, \bibinfo {author} {\bibfnamefont {L.}~\bibnamefont {Radzihovsky}}, \bibinfo {author} {\bibfnamefont {M.~A.}\ \bibnamefont {Glaser}}, \bibinfo {author} {\bibfnamefont {J.~E.}\ \bibnamefont {Maclennan}}, \bibinfo {author} {\bibfnamefont {D.}~\bibnamefont {Bedrov}}, \bibinfo {author} {\bibfnamefont {D.~M.}\ \bibnamefont {Walba}}, \emph {et~al.},\ }\bibfield  {title} {\bibinfo {title} {First-principles experimental demonstration of ferroelectricity in a thermotropic nematic liquid crystal: Polar domains and striking electro-optics},\ }\href@noop {} {\bibfield  {journal} {\bibinfo  {journal} {Proceedings of the National Academy of Sciences}\ }\textbf {\bibinfo {volume} {117}},\ \bibinfo {pages}
  {14021} (\bibinfo {year} {2020})}\BibitemShut {NoStop}%
\bibitem [{\citenamefont {Lavrentovich}(2020)}]{lavrentovich2020ferroelectric}%
  \BibitemOpen
  \bibfield  {author} {\bibinfo {author} {\bibfnamefont {O.~D.}\ \bibnamefont {Lavrentovich}},\ }\bibfield  {title} {\bibinfo {title} {Ferroelectric nematic liquid crystal, a century in waiting},\ }\href@noop {} {\bibfield  {journal} {\bibinfo  {journal} {Proceedings of the National Academy of Sciences}\ }\textbf {\bibinfo {volume} {117}},\ \bibinfo {pages} {14629} (\bibinfo {year} {2020})}\BibitemShut {NoStop}%
\bibitem [{\citenamefont {Sebasti{\'a}n}\ \emph {et~al.}(2022)\citenamefont {Sebasti{\'a}n}, \citenamefont {{\v{C}}opi{\v{c}}},\ and\ \citenamefont {Mertelj}}]{sebastian2022ferroelectric}%
  \BibitemOpen
  \bibfield  {author} {\bibinfo {author} {\bibfnamefont {N.}~\bibnamefont {Sebasti{\'a}n}}, \bibinfo {author} {\bibfnamefont {M.}~\bibnamefont {{\v{C}}opi{\v{c}}}},\ and\ \bibinfo {author} {\bibfnamefont {A.}~\bibnamefont {Mertelj}},\ }\bibfield  {title} {\bibinfo {title} {Ferroelectric nematic liquid-crystalline phases},\ }\href@noop {} {\bibfield  {journal} {\bibinfo  {journal} {Physical Review E}\ }\textbf {\bibinfo {volume} {106}},\ \bibinfo {pages} {021001} (\bibinfo {year} {2022})}\BibitemShut {NoStop}%
\bibitem [{\citenamefont {Eckert}\ \emph {et~al.}(2023)\citenamefont {Eckert}, \citenamefont {Ladoux}, \citenamefont {M{\`e}ge}, \citenamefont {Giomi},\ and\ \citenamefont {Schmidt}}]{eckert2023hexanematic}%
  \BibitemOpen
  \bibfield  {author} {\bibinfo {author} {\bibfnamefont {J.}~\bibnamefont {Eckert}}, \bibinfo {author} {\bibfnamefont {B.}~\bibnamefont {Ladoux}}, \bibinfo {author} {\bibfnamefont {R.-M.}\ \bibnamefont {M{\`e}ge}}, \bibinfo {author} {\bibfnamefont {L.}~\bibnamefont {Giomi}},\ and\ \bibinfo {author} {\bibfnamefont {T.}~\bibnamefont {Schmidt}},\ }\bibfield  {title} {\bibinfo {title} {Hexanematic crossover in epithelial monolayers depends on cell adhesion and cell density},\ }\href@noop {} {\bibfield  {journal} {\bibinfo  {journal} {Nature Communications}\ }\textbf {\bibinfo {volume} {14}},\ \bibinfo {pages} {5762} (\bibinfo {year} {2023})}\BibitemShut {NoStop}%
\bibitem [{\citenamefont {Armengol-Collado}\ \emph {et~al.}(2023)\citenamefont {Armengol-Collado}, \citenamefont {Carenza}, \citenamefont {Eckert}, \citenamefont {Krommydas},\ and\ \citenamefont {Giomi}}]{armengol2023epithelia}%
  \BibitemOpen
  \bibfield  {author} {\bibinfo {author} {\bibfnamefont {J.-M.}\ \bibnamefont {Armengol-Collado}}, \bibinfo {author} {\bibfnamefont {L.~N.}\ \bibnamefont {Carenza}}, \bibinfo {author} {\bibfnamefont {J.}~\bibnamefont {Eckert}}, \bibinfo {author} {\bibfnamefont {D.}~\bibnamefont {Krommydas}},\ and\ \bibinfo {author} {\bibfnamefont {L.}~\bibnamefont {Giomi}},\ }\bibfield  {title} {\bibinfo {title} {Epithelia are multiscale active liquid crystals},\ }\href@noop {} {\bibfield  {journal} {\bibinfo  {journal} {Nature Physics}\ }\textbf {\bibinfo {volume} {19}},\ \bibinfo {pages} {1773} (\bibinfo {year} {2023})}\BibitemShut {NoStop}%
\bibitem [{\citenamefont {Armengol-Collado}\ \emph {et~al.}(2024)\citenamefont {Armengol-Collado}, \citenamefont {Carenza},\ and\ \citenamefont {Giomi}}]{armengol2024hydrodynamics}%
  \BibitemOpen
  \bibfield  {author} {\bibinfo {author} {\bibfnamefont {J.-M.}\ \bibnamefont {Armengol-Collado}}, \bibinfo {author} {\bibfnamefont {L.~N.}\ \bibnamefont {Carenza}},\ and\ \bibinfo {author} {\bibfnamefont {L.}~\bibnamefont {Giomi}},\ }\bibfield  {title} {\bibinfo {title} {Hydrodynamics and multiscale orderin confluent epithelia},\ }\href@noop {} {\bibfield  {journal} {\bibinfo  {journal} {Elife}\ }\textbf {\bibinfo {volume} {13}},\ \bibinfo {pages} {e86400} (\bibinfo {year} {2024})}\BibitemShut {NoStop}%
\bibitem [{\citenamefont {Bruinsma}\ and\ \citenamefont {Aeppli}(1982)}]{bruinsma1982hexatic}%
  \BibitemOpen
  \bibfield  {author} {\bibinfo {author} {\bibfnamefont {R.}~\bibnamefont {Bruinsma}}\ and\ \bibinfo {author} {\bibfnamefont {G.}~\bibnamefont {Aeppli}},\ }\bibfield  {title} {\bibinfo {title} {Hexatic order and herring-bone packing in liquid crystals},\ }\href@noop {} {\bibfield  {journal} {\bibinfo  {journal} {Physical Review Letters}\ }\textbf {\bibinfo {volume} {48}},\ \bibinfo {pages} {1625} (\bibinfo {year} {1982})}\BibitemShut {NoStop}%
\bibitem [{\citenamefont {Dierker}\ and\ \citenamefont {Pindak}(1987)}]{dierker1987dynamics}%
  \BibitemOpen
  \bibfield  {author} {\bibinfo {author} {\bibfnamefont {S.}~\bibnamefont {Dierker}}\ and\ \bibinfo {author} {\bibfnamefont {R.}~\bibnamefont {Pindak}},\ }\bibfield  {title} {\bibinfo {title} {Dynamics of thin tilted hexatic liquid crystal films},\ }\href@noop {} {\bibfield  {journal} {\bibinfo  {journal} {Physical review letters}\ }\textbf {\bibinfo {volume} {59}},\ \bibinfo {pages} {1002} (\bibinfo {year} {1987})}\BibitemShut {NoStop}%
\bibitem [{\citenamefont {Sprunt}\ and\ \citenamefont {Litster}(1987)}]{sprunt1987light}%
  \BibitemOpen
  \bibfield  {author} {\bibinfo {author} {\bibfnamefont {S.}~\bibnamefont {Sprunt}}\ and\ \bibinfo {author} {\bibfnamefont {J.}~\bibnamefont {Litster}},\ }\bibfield  {title} {\bibinfo {title} {Light-scattering study of bond orientational order in a tilted hexatic liquid-crystal film},\ }\href@noop {} {\bibfield  {journal} {\bibinfo  {journal} {Physical review letters}\ }\textbf {\bibinfo {volume} {59}},\ \bibinfo {pages} {2682} (\bibinfo {year} {1987})}\BibitemShut {NoStop}%
\bibitem [{\citenamefont {Selinger}\ and\ \citenamefont {Nelson}(1989)}]{selinger1989theory}%
  \BibitemOpen
  \bibfield  {author} {\bibinfo {author} {\bibfnamefont {J.~V.}\ \bibnamefont {Selinger}}\ and\ \bibinfo {author} {\bibfnamefont {D.~R.}\ \bibnamefont {Nelson}},\ }\bibfield  {title} {\bibinfo {title} {Theory of transitions among tilted hexatic phases in liquid crystals},\ }\href@noop {} {\bibfield  {journal} {\bibinfo  {journal} {Physical Review A}\ }\textbf {\bibinfo {volume} {39}},\ \bibinfo {pages} {3135} (\bibinfo {year} {1989})}\BibitemShut {NoStop}%
\bibitem [{\citenamefont {Selinger}(1991)}]{selinger1991dynamics}%
  \BibitemOpen
  \bibfield  {author} {\bibinfo {author} {\bibfnamefont {J.~V.}\ \bibnamefont {Selinger}},\ }\bibfield  {title} {\bibinfo {title} {Dynamics of tilted hexatic phases in liquid-crystal films},\ }\href@noop {} {\bibfield  {journal} {\bibinfo  {journal} {Journal de Physique II}\ }\textbf {\bibinfo {volume} {1}},\ \bibinfo {pages} {1363} (\bibinfo {year} {1991})}\BibitemShut {NoStop}%
\bibitem [{\citenamefont {Drouin-Touchette}\ \emph {et~al.}(2022)\citenamefont {Drouin-Touchette}, \citenamefont {Orth}, \citenamefont {Coleman}, \citenamefont {Chandra},\ and\ \citenamefont {Lubensky}}]{drouin2022emergent}%
  \BibitemOpen
  \bibfield  {author} {\bibinfo {author} {\bibfnamefont {V.}~\bibnamefont {Drouin-Touchette}}, \bibinfo {author} {\bibfnamefont {P.~P.}\ \bibnamefont {Orth}}, \bibinfo {author} {\bibfnamefont {P.}~\bibnamefont {Coleman}}, \bibinfo {author} {\bibfnamefont {P.}~\bibnamefont {Chandra}},\ and\ \bibinfo {author} {\bibfnamefont {T.~C.}\ \bibnamefont {Lubensky}},\ }\bibfield  {title} {\bibinfo {title} {Emergent potts order in a coupled hexatic-nematic xy model},\ }\href@noop {} {\bibfield  {journal} {\bibinfo  {journal} {Physical Review X}\ }\textbf {\bibinfo {volume} {12}},\ \bibinfo {pages} {011043} (\bibinfo {year} {2022})}\BibitemShut {NoStop}%
\bibitem [{\citenamefont {Anisimov}\ \emph {et~al.}(1981)\citenamefont {Anisimov}, \citenamefont {Gorodetski{\u\i}},\ and\ \citenamefont {Zaprudski{\u\i}}}]{anisimov1981phase}%
  \BibitemOpen
  \bibfield  {author} {\bibinfo {author} {\bibfnamefont {M.}~\bibnamefont {Anisimov}}, \bibinfo {author} {\bibfnamefont {E.}~\bibnamefont {Gorodetski{\u\i}}},\ and\ \bibinfo {author} {\bibfnamefont {V.}~\bibnamefont {Zaprudski{\u\i}}},\ }\bibfield  {title} {\bibinfo {title} {Phase transitions with coupled order parameters},\ }\href@noop {} {\bibfield  {journal} {\bibinfo  {journal} {Soviet Physics Uspekhi}\ }\textbf {\bibinfo {volume} {24}},\ \bibinfo {pages} {57} (\bibinfo {year} {1981})}\BibitemShut {NoStop}%
\bibitem [{\citenamefont {Meyer}(1969)}]{meyer1969piezoelectric}%
  \BibitemOpen
  \bibfield  {author} {\bibinfo {author} {\bibfnamefont {R.~B.}\ \bibnamefont {Meyer}},\ }\bibfield  {title} {\bibinfo {title} {Piezoelectric effects in liquid crystals},\ }\href@noop {} {\bibfield  {journal} {\bibinfo  {journal} {Physical Review Letters}\ }\textbf {\bibinfo {volume} {22}},\ \bibinfo {pages} {918} (\bibinfo {year} {1969})}\BibitemShut {NoStop}%
\bibitem [{\citenamefont {Shamid}\ \emph {et~al.}(2013)\citenamefont {Shamid}, \citenamefont {Dhakal},\ and\ \citenamefont {Selinger}}]{shamid2013statistical}%
  \BibitemOpen
  \bibfield  {author} {\bibinfo {author} {\bibfnamefont {S.~M.}\ \bibnamefont {Shamid}}, \bibinfo {author} {\bibfnamefont {S.}~\bibnamefont {Dhakal}},\ and\ \bibinfo {author} {\bibfnamefont {J.~V.}\ \bibnamefont {Selinger}},\ }\bibfield  {title} {\bibinfo {title} {Statistical mechanics of bend flexoelectricity and the twist-bend phase in bent-core liquid crystals},\ }\href@noop {} {\bibfield  {journal} {\bibinfo  {journal} {Physical Review E—Statistical, Nonlinear, and Soft Matter Physics}\ }\textbf {\bibinfo {volume} {87}},\ \bibinfo {pages} {052503} (\bibinfo {year} {2013})}\BibitemShut {NoStop}%
\bibitem [{\citenamefont {Mertelj}\ \emph {et~al.}(2018)\citenamefont {Mertelj}, \citenamefont {Cmok}, \citenamefont {Sebasti{\'a}n}, \citenamefont {Mandle}, \citenamefont {Parker}, \citenamefont {Whitwood}, \citenamefont {Goodby},\ and\ \citenamefont {{\v{C}}opi{\v{c}}}}]{mertelj2018splay}%
  \BibitemOpen
  \bibfield  {author} {\bibinfo {author} {\bibfnamefont {A.}~\bibnamefont {Mertelj}}, \bibinfo {author} {\bibfnamefont {L.}~\bibnamefont {Cmok}}, \bibinfo {author} {\bibfnamefont {N.}~\bibnamefont {Sebasti{\'a}n}}, \bibinfo {author} {\bibfnamefont {R.~J.}\ \bibnamefont {Mandle}}, \bibinfo {author} {\bibfnamefont {R.~R.}\ \bibnamefont {Parker}}, \bibinfo {author} {\bibfnamefont {A.~C.}\ \bibnamefont {Whitwood}}, \bibinfo {author} {\bibfnamefont {J.~W.}\ \bibnamefont {Goodby}},\ and\ \bibinfo {author} {\bibfnamefont {M.}~\bibnamefont {{\v{C}}opi{\v{c}}}},\ }\bibfield  {title} {\bibinfo {title} {Splay nematic phase},\ }\href@noop {} {\bibfield  {journal} {\bibinfo  {journal} {Physical Review X}\ }\textbf {\bibinfo {volume} {8}},\ \bibinfo {pages} {041025} (\bibinfo {year} {2018})}\BibitemShut {NoStop}%
\bibitem [{\citenamefont {Sebasti{\'a}n}\ \emph {et~al.}(2020)\citenamefont {Sebasti{\'a}n}, \citenamefont {Cmok}, \citenamefont {Mandle}, \citenamefont {de~la Fuente}, \citenamefont {Dreven{\v{s}}ek~Olenik}, \citenamefont {{\v{C}}opi{\v{c}}},\ and\ \citenamefont {Mertelj}}]{sebastian2020ferroelectric}%
  \BibitemOpen
  \bibfield  {author} {\bibinfo {author} {\bibfnamefont {N.}~\bibnamefont {Sebasti{\'a}n}}, \bibinfo {author} {\bibfnamefont {L.}~\bibnamefont {Cmok}}, \bibinfo {author} {\bibfnamefont {R.~J.}\ \bibnamefont {Mandle}}, \bibinfo {author} {\bibfnamefont {M.~R.}\ \bibnamefont {de~la Fuente}}, \bibinfo {author} {\bibfnamefont {I.}~\bibnamefont {Dreven{\v{s}}ek~Olenik}}, \bibinfo {author} {\bibfnamefont {M.}~\bibnamefont {{\v{C}}opi{\v{c}}}},\ and\ \bibinfo {author} {\bibfnamefont {A.}~\bibnamefont {Mertelj}},\ }\bibfield  {title} {\bibinfo {title} {Ferroelectric-ferroelastic phase transition in a nematic liquid crystal},\ }\href@noop {} {\bibfield  {journal} {\bibinfo  {journal} {Physical review letters}\ }\textbf {\bibinfo {volume} {124}},\ \bibinfo {pages} {037801} (\bibinfo {year} {2020})}\BibitemShut {NoStop}%
\bibitem [{\citenamefont {Rosseto}\ and\ \citenamefont {Selinger}(2020)}]{rosseto2020theory}%
  \BibitemOpen
  \bibfield  {author} {\bibinfo {author} {\bibfnamefont {M.~P.}\ \bibnamefont {Rosseto}}\ and\ \bibinfo {author} {\bibfnamefont {J.~V.}\ \bibnamefont {Selinger}},\ }\bibfield  {title} {\bibinfo {title} {Theory of the splay nematic phase: single versus double splay},\ }\href@noop {} {\bibfield  {journal} {\bibinfo  {journal} {Physical Review E}\ }\textbf {\bibinfo {volume} {101}},\ \bibinfo {pages} {052707} (\bibinfo {year} {2020})}\BibitemShut {NoStop}%
\bibitem [{\citenamefont {Paik}\ and\ \citenamefont {Selinger}(2024)}]{paik2024flexoelectricity}%
  \BibitemOpen
  \bibfield  {author} {\bibinfo {author} {\bibfnamefont {L.}~\bibnamefont {Paik}}\ and\ \bibinfo {author} {\bibfnamefont {J.~V.}\ \bibnamefont {Selinger}},\ }\bibfield  {title} {\bibinfo {title} {Flexoelectricity versus electrostatics in polar nematic liquid crystals},\ }\href@noop {} {\bibfield  {journal} {\bibinfo  {journal} {arXiv preprint arXiv:2408.10347}\ } (\bibinfo {year} {2024})}\BibitemShut {NoStop}%
\bibitem [{\citenamefont {Korshunov}(1985)}]{korshunov1985possible}%
  \BibitemOpen
  \bibfield  {author} {\bibinfo {author} {\bibfnamefont {S.~E.}\ \bibnamefont {Korshunov}},\ }\bibfield  {title} {\bibinfo {title} {Possible splitting of a phase transition in a $2d$ $xy$ model},\ }\href@noop {} {\bibfield  {journal} {\bibinfo  {journal} {JETP Lett.}\ }\textbf {\bibinfo {volume} {41}},\ \bibinfo {pages} {216} (\bibinfo {year} {1985})}\BibitemShut {NoStop}%
\bibitem [{\citenamefont {Lee}\ and\ \citenamefont {Grinstein}(1985)}]{lee1985strings}%
  \BibitemOpen
  \bibfield  {author} {\bibinfo {author} {\bibfnamefont {D.}~\bibnamefont {Lee}}\ and\ \bibinfo {author} {\bibfnamefont {G.}~\bibnamefont {Grinstein}},\ }\bibfield  {title} {\bibinfo {title} {Strings in two-dimensional classical xy models},\ }\href@noop {} {\bibfield  {journal} {\bibinfo  {journal} {Physical review letters}\ }\textbf {\bibinfo {volume} {55}},\ \bibinfo {pages} {541} (\bibinfo {year} {1985})}\BibitemShut {NoStop}%
\bibitem [{\citenamefont {Carpenter}\ and\ \citenamefont {Chalker}(1989)}]{carpenter1989phase}%
  \BibitemOpen
  \bibfield  {author} {\bibinfo {author} {\bibfnamefont {D.}~\bibnamefont {Carpenter}}\ and\ \bibinfo {author} {\bibfnamefont {J.}~\bibnamefont {Chalker}},\ }\bibfield  {title} {\bibinfo {title} {The phase diagram of a generalised $xy$ model},\ }\href@noop {} {\bibfield  {journal} {\bibinfo  {journal} {Journal of Physics: Condensed Matter}\ }\textbf {\bibinfo {volume} {1}},\ \bibinfo {pages} {4907} (\bibinfo {year} {1989})}\BibitemShut {NoStop}%
\bibitem [{\citenamefont {Shi}\ \emph {et~al.}(2011)\citenamefont {Shi}, \citenamefont {Lamacraft},\ and\ \citenamefont {Fendley}}]{shi2011boson}%
  \BibitemOpen
  \bibfield  {author} {\bibinfo {author} {\bibfnamefont {Y.}~\bibnamefont {Shi}}, \bibinfo {author} {\bibfnamefont {A.}~\bibnamefont {Lamacraft}},\ and\ \bibinfo {author} {\bibfnamefont {P.}~\bibnamefont {Fendley}},\ }\bibfield  {title} {\bibinfo {title} {Boson pairing and unusual criticality in a generalized xy model},\ }\href@noop {} {\bibfield  {journal} {\bibinfo  {journal} {Physical review letters}\ }\textbf {\bibinfo {volume} {107}},\ \bibinfo {pages} {240601} (\bibinfo {year} {2011})}\BibitemShut {NoStop}%
\bibitem [{\citenamefont {Amiri}\ \emph {et~al.}(2022)\citenamefont {Amiri}, \citenamefont {Mueller},\ and\ \citenamefont {Doostmohammadi}}]{amiri2022unifying}%
  \BibitemOpen
  \bibfield  {author} {\bibinfo {author} {\bibfnamefont {A.}~\bibnamefont {Amiri}}, \bibinfo {author} {\bibfnamefont {R.}~\bibnamefont {Mueller}},\ and\ \bibinfo {author} {\bibfnamefont {A.}~\bibnamefont {Doostmohammadi}},\ }\bibfield  {title} {\bibinfo {title} {Unifying polar and nematic active matter: emergence and co-existence of half-integer and full-integer topological defects},\ }\href@noop {} {\bibfield  {journal} {\bibinfo  {journal} {Journal of Physics A: Mathematical and Theoretical}\ }\textbf {\bibinfo {volume} {55}},\ \bibinfo {pages} {094002} (\bibinfo {year} {2022})}\BibitemShut {NoStop}%
\bibitem [{\citenamefont {Ruider}\ \emph {et~al.}(2024)\citenamefont {Ruider}, \citenamefont {Thijssen}, \citenamefont {Vannier}, \citenamefont {Paloschi}, \citenamefont {Sciortino}, \citenamefont {Doostmohammadi},\ and\ \citenamefont {Bausch}}]{ruider2024topological}%
  \BibitemOpen
  \bibfield  {author} {\bibinfo {author} {\bibfnamefont {I.}~\bibnamefont {Ruider}}, \bibinfo {author} {\bibfnamefont {K.}~\bibnamefont {Thijssen}}, \bibinfo {author} {\bibfnamefont {D.~R.}\ \bibnamefont {Vannier}}, \bibinfo {author} {\bibfnamefont {V.}~\bibnamefont {Paloschi}}, \bibinfo {author} {\bibfnamefont {A.}~\bibnamefont {Sciortino}}, \bibinfo {author} {\bibfnamefont {A.}~\bibnamefont {Doostmohammadi}},\ and\ \bibinfo {author} {\bibfnamefont {A.~R.}\ \bibnamefont {Bausch}},\ }\bibfield  {title} {\bibinfo {title} {Topological excitations govern ordering kinetics in endothelial cell layers},\ }\bibfield  {journal} {\bibinfo  {journal} {bioRxiv}\ }\href {https://doi.org/10.1101/2024.09.26.615134} {10.1101/2024.09.26.615134} (\bibinfo {year} {2024}),\ \Eprint {https://arxiv.org/abs/https://www.biorxiv.org/content/early/2024/09/26/2024.09.26.615134.full.pdf} {https://www.biorxiv.org/content/early/2024/09/26/2024.09.26.615134.full.pdf} \BibitemShut {NoStop}%
\bibitem [{\citenamefont {Schiesser}(2012)}]{schiesser2012numerical}%
  \BibitemOpen
  \bibfield  {author} {\bibinfo {author} {\bibfnamefont {W.~E.}\ \bibnamefont {Schiesser}},\ }\href@noop {} {\emph {\bibinfo {title} {The numerical method of lines: integration of partial differential equations}}}\ (\bibinfo  {publisher} {Elsevier},\ \bibinfo {year} {2012})\BibitemShut {NoStop}%
\bibitem [{\citenamefont {Press}\ \emph {et~al.}(1992)\citenamefont {Press}, \citenamefont {Flannery}, \citenamefont {Teukolsky},\ and\ \citenamefont {Vetterling}}]{press1992multistep}%
  \BibitemOpen
  \bibfield  {author} {\bibinfo {author} {\bibfnamefont {W.}~\bibnamefont {Press}}, \bibinfo {author} {\bibfnamefont {B.}~\bibnamefont {Flannery}}, \bibinfo {author} {\bibfnamefont {S.}~\bibnamefont {Teukolsky}},\ and\ \bibinfo {author} {\bibfnamefont {W.}~\bibnamefont {Vetterling}},\ }\bibfield  {title} {\bibinfo {title} {Multistep, multivalue, and predictor-corrector methods},\ }\href@noop {} {\bibfield  {journal} {\bibinfo  {journal} {Numerical Recipes in FORTRAN: The Art of Scientific Computing}\ ,\ \bibinfo {pages} {740}} (\bibinfo {year} {1992})}\BibitemShut {NoStop}%
\end{thebibliography}%

    	%%%%%%%%%% Merge with supplemental materials %%%%%%%%%%
	\onecolumngrid
	\clearpage
	\begin{center}
		\textbf{\large Supplementary Material}
	\end{center}
	\appendix
	
	\setcounter{figure}{0}
	\setcounter{section}{0}
	\setcounter{equation}{0}
	\renewcommand{\thefigure}{S\arabic{figure}}
	\renewcommand{\theequation}{S.\arabic{equation}}

    \section{Phase diagram}
    \label{app:phase_diagram}
    
    Here we minimize $\mathcal F$, given by
    \beq \mathcal F = \mathcal A_{NP} (Q-p^2)^2 + \mathcal A_P p^2 + (Q^2-1)^2 \eeq
    
	Setting derivatives of $\mathcal F$ to zero, we have
	\begin{align}
		\p{\mathcal F}{p} &= 0 \implies -2\mathcal A_{NP} p(Q - p^2) + \mathcal A_P p = 0 \\
		\p{\mathcal F}{Q} &= 0 \implies \mathcal A_{NP} (Q - p^2) + 2 Q(Q^2 - 1) = 0
	\end{align}
	
    %From the first equation, if $p \neq 0$, then the $Q$ order is locked to the $p$ order (i.e. $Q$ can be explicitly written in terms of $p$). This means that in this case the $p$-atic order parameter is reduced to the $k$-atic order, thus reducing the symmetry and prohibiting $p$-atic defects.

    By inspection, $p=0$ is a solution when $Q = 0$ or $Q = \sqrt{1 - \frac{\mathcal A_{NP}}{2}}$ for $\mathcal A_{NP} \le 2$. We now determine which solution gives lower energy for $\mathcal A_{NP} < 2$. Substituting into $\mathcal F$, we find
    
	\begin{align}
		\mathcal F(p=0, Q=0) &= 1 \\
		\mathcal F\left(p=0, Q = \sqrt{1 - \mathcal A_{NP}/2}\right) &= -\frac{1}{4} (\mathcal A_{NP} -4) \mathcal A_{NP} \label{eq:F_nematic}
	\end{align}
	By inspection, for $\mathcal A_{NP} \ge 2$, $1 \ge -\frac{1}{4} (\mathcal A_{NP} -4) \mathcal A_{NP}$, with equality for $\mathcal A_{NP}=2$. Thus now the possible solutions are
	\begin{enumerate}
	\item $p = 0$ and $Q = 0$ for $\mathcal A_{NP} \ge 2$
	\item $p = 0$ and $Q = \sqrt{1 - \frac{\mathcal A_{NP}}{2}}$ for $\mathcal A_{NP} \le 2$.
\end{enumerate}

We now look for non-zero $p$ solution. We have
	\begin{align}
	\p{\mathcal F}{p} &= 0 \implies -2\mathcal A_{NP} (Q - p^2) + \mathcal A_P = 0 \\
	\p{\mathcal F}{Q} &= 0 \implies \mathcal A_{NP} (Q - p^2) + 2 Q(Q^2 - 1) = 0
\end{align}

From the first equation, we learn that
\beq Q = p^2 + \frac{\mathcal A_P}{2\mathcal A_{NP}} \label{eq:Q1soln}\eeq
%\beq p = \sqrt{Q - \frac{\mathcal A_P}{2\mathcal A_{NP}}} \eeq
Thus the nematic order $Q$ is locked to the polar order $p$, and when this is the case, the nematic order can be written in terms of polar order, which means that half-integer defects are no longer allowed. We will keep in mind that in order for this solution to be valid, we must have $Q > \mathcal A_P/(2\mathcal A_{NP})$.

Now substituting for $p$ in the 2nd equation gives
\beq Q^3 - Q + \frac{\mathcal A_P}{4} = 0\label{eq:Q-cubic}.\eeq
Since $\mathcal A_P/4 > 0$, by Vieta's formula, provided that the discriminant $\Delta = -(27/16)(\mathcal A_P^2 - 64/27) \ge 0$, i.e., $\mathcal A_P \le 8/\sqrt{27}$, the only positive solutions for $Q$ are
\beq Q = 2 \, \re{\sqrt[3]{-\frac{\mathcal A_P}{8} + i\sqrt{|D|}}}, \quad 2\, \re{e^{4\pi i/3}\sqrt[3]{-\frac{\mathcal A_P}{8} + i\sqrt{|D|}}} , \eeq
where $D = -\Delta / (4 \times 27)$.

It can be shown that $\mathcal F$ is smaller when the first solution is taken. This solution is valid when Eq.~\eqref{eq:Q1soln} is satisfied, which in terms of $\mathcal A_P$ and $\mathcal A_{NP}$, becomes
\beq \mathcal A_{NP} \ge \frac{\mathcal A_P}{2 Q} = \mathcal A_P/\left(4 \, \re{\sqrt[3]{-\frac{\mathcal A_P}{8} + i\sqrt{|D|}}}\right)\eeq
 Thus to summarize, the solution is
	\beq Q = 2 \, \re{\sqrt[3]{-\frac{\mathcal A_P}{8} + i\sqrt{|D|}}} \label{eq:Qsoln}\eeq
	with $D = -1/27 + \mathcal A_P^2/64$, provided that $\mathcal A_P \le 8/\sqrt{27}$ and $Q \ge \mathcal A_P/(2\mathcal A_{NP})$.

To summarize what we have learned so far, we have three phases:
\begin{enumerate}
	\item Isotropic phase: $p = 0$ and $Q = 0$ for $\mathcal A_{NP} \ge 2$.
	\item Nematic phase: $p = 0$ and $Q = \sqrt{1 - \mathcal A_{NP}/2}$ for $\mathcal A_{NP} \le 2$.
	\item Nematopolar phase: $p = \sqrt{Q - \mathcal A_P/(2\mathcal A_{NP})}$ and $Q = 2 \, \re{\sqrt[3]{-\frac{\mathcal A_P}{8} + i\sqrt{|D|}}}$, with $D = -1/27 + \mathcal A_P^2/64$, provided that $\mathcal A_P \le 8/\sqrt{27}$ and $Q \ge \mathcal A_P/(2\mathcal A_{NP})$.
\end{enumerate}

Now that we have identified the phases, we find the boundaries. We first determine the boundary between the nematic and nematopolar phases. The phase boundary is the curve $\mathcal A_P(\mathcal A_{NP})$ for $\mathcal A_{NP} \le 2$ such that
\beq \mathcal F\left(p=0, Q = \sqrt{1 - \mathcal A_{NP}/2}\right) = \mathcal F\left(p = \sqrt{Q - \mathcal A_P/(2\mathcal A_{NP})}, Q = 2 \, \re{\sqrt[3]{-\frac{\mathcal A_P}{8} + i\sqrt{|D|}}}\right)\label{eq:phase_boundary_small_lambda}\eeq
where $D = -1/27 + \mathcal A_P^2/64$. The phase boundary is continuous if the fields $p$ and $Q$ match in both phases, and discontinuous otherwise. The phase boundary is continuous if $p=0$, which implies $Q = \mathcal A_P/(2\mathcal A_{NP}) = \sqrt{1 - \mathcal A_{NP}/2}$, leading to
\beq \mathcal A_P(\mathcal A_{NP}) = 2\mathcal A_{NP} \sqrt{1-\mathcal A_{NP}/2}\eeq
Note that since $\mathcal A_P(\mathcal A_{NP} = 4/3) = 8/\sqrt{27}$, the continuous part of the phase boundary has endpoints at $(\mathcal A_P, \mathcal A_{NP}) = (0,0)$ and $(\mathcal A_P, \mathcal A_{NP}) = (8/\sqrt{27},4/3)$. Hence the point $(\mathcal A_P, \mathcal A_{NP}) = (8/\sqrt{27},4/3)$ is a tricritical point, where a continuous and a discontinuous phase transition meet.

We now need to determine the continuation of the boundary from $\mathcal A_{NP} = 4/3$ to $\mathcal A_{NP} = 2$. We first note that
\beq \mathcal F\left(p = \sqrt{Q - \mathcal A_P/(2\mathcal A_{NP})}, Q\right)  = -\frac{1}{4} \frac{\mathcal A_P^2}{\mathcal A_{NP}} + \mathcal A_P Q + (Q^2  - 1)^2 .\label{eq:F_locked}\eeq
Substituting Eqs. \eqref{eq:F_nematic} and \eqref{eq:F_locked} into
Eq.~\eqref{eq:phase_boundary_small_lambda}, we find that
\beq -\frac{1}{4}\mathcal A_{NP}(\mathcal A_{NP}-4) =  -\frac{1}{4} \frac{\mathcal A_P^2}{\mathcal A_{NP}} + \mathcal A_P Q + (Q^2  - 1)^2 .\eeq
Upon multiplication by $4\mathcal A_{NP}$ and rearrangement, we arrive at
\beq \mathcal A_{NP}^3-4\mathcal A_{NP}^2 + \left[4 \mathcal A_P Q + 4(Q^2  - 1)^2\right]\mathcal A_{NP} - \mathcal A_P^2  = 0 .\eeq
Recalling that $Q$ is independent of $\mathcal A_{NP}$, then we have a cubic equation for $\mathcal A_{NP}$, which we now solve. 

Upon the following change of variables from $\mathcal A_{NP}$ to $\tilde{\mathcal A}_{NP}$ given by
\beq\mathcal A_{NP} = \tilde{\mathcal A}_{NP} + 4/3 ,\eeq
we now need to solve
\beq \tilde{\mathcal A}_{NP} ^3 + a \tilde{\mathcal A}_{NP} + b = 0 \eeq
where
\begin{align}
	a &= 4\left(\mathcal A_P Q + (Q^2  - 1)^2\right) - \frac{16}{3} \\
	b &= -\frac{128}{27} + \frac{16}{3}\left(\mathcal A_P Q + (Q^2  - 1)^2\right) -\mathcal A_P^2 ,
\end{align}
and $Q$ is given by Eq.~\eqref{eq:Qsoln}.

We note that since we can express the discriminant $\Delta$ as
\beq \Delta = -(4 \times 27)\left(Q^3 - Q + \frac{\mathcal A_P}{4}\right)\left(-\frac{64}{27}Q^3 + \frac{64}{27}Q - \frac{32}{27}\mathcal A_P + \frac{16}{3}\mathcal A_P Q^2 - 4\mathcal A_P^2 Q + \mathcal A_P^3\right), \eeq
then since $Q$ satisfies Eq.~\eqref{eq:Q-cubic}, $\Delta=0$. Hence,
\beq \mathcal A_{NP} = \tilde{\mathcal A}_{NP} + 4/3 = 2\left(-\frac{b}{2}\right)^{1/3} + 4/3 \label{eq:discontinuous_lower}\eeq
In the limit $\mathcal A_{NP} \gg 1$, we have $p = 0$, and hence the boundary for large $\mathcal A_{NP}$ asymptotes to $\mathcal A_P =  4\sqrt{6}/9$, which is obtained by solving the systems of equations
\begin{align}
    \mathcal F\left(p = 0, Q\right) &= 1\\
    \left.\p{\mathcal F}{Q}\right|_{p=0} &= 0
\end{align}
for $Q$ and $\mathcal A_P$.

We now determine the phase boundary between the nematopolar and isotropic phases. Since in the nematopolar phase, we cannot have both $p$ and $Q$ vanish, then the phase boundary between the nematopolar and isotropic phases is a discontinuous phase transition. The phase boundary is determined by
\beq \mathcal F(p = 0, Q = 0) = \mathcal F\left(p = \sqrt{Q - \mathcal A_P/(2\mathcal A_{NP})}, Q = 2 \, \re{\sqrt[3]{-\frac{\mathcal A_P}{8} + i\sqrt{|D|}}}\right),\eeq
where $D = -1/27 + \mathcal A_P^2/64$. Recalling that $\mathcal F(0,0) = 1$, and using Eq.~\eqref{eq:F_locked}, we have
\beq 1 = -\frac{1}{4} \frac{\mathcal A_P^2}{\mathcal A_{NP}} + \mathcal A_P Q + (Q^2  - 1)^2 \eeq
and hence the phase boundary is determined by
\beq \mathcal A_{NP}  = \frac{\mathcal A_P^2}{4(\mathcal A_P Q + (Q^2  - 1)^2 - 1)} \label{eq:discontinous_upper},\eeq
where $Q$ satisfies Eq.~\eqref{eq:Qsoln}.

Finally, the triple point, where all three phases meet, is at $(\mathcal A_P, \mathcal A_{NP}) = (\sqrt{2}, 2)$, which is when
\beq \mathcal F(p = 0, Q = 0) = \mathcal F(p = 0, Q = \sqrt{1 - \mathcal A_{NP}/2} = \mathcal F\left(p = \sqrt{Q - \mathcal A_P/(2\mathcal A_{NP})}, Q = 2 \, \re{\sqrt[3]{-\frac{\mathcal A_P}{8} + i\sqrt{|D|}}}\right),\eeq
where $D = -1/27 + \mathcal A_P^2/64$.

    \section{Related model}
    \label{app:polarnematic}

A related model, the polarnematic model, where the roles of $\vec p$ and $\overleftrightarrow{Q}$ are reversed, that is, $\vec p$ is ordered whereas $\overleftrightarrow{Q}$ is isotropic, has the following free energy:
\beq \mathcal F = \mathcal A_{NP} \left|\overleftrightarrow{Q} - \overleftrightarrow{P}\right|^2 + \mathcal A_P |Q|^2 + (|p|^2 - 1)^2 . \eeq
$\mathcal F$ is minimized when $\overleftrightarrow{Q}$ and $\overleftrightarrow{P}$ are aligned. In terms of the magnitudes $Q$ and $p$ of $\overleftrightarrow{Q}$ and $\vec p$, respectively, $\mathcal F$ becomes
\beq \mathcal F = \mathcal A_{NP} (Q - p^2)^2 + \mathcal A_P Q^2 + (p^2 - 1)^2 , \eeq
which is minimized when
\begin{align}
	p &= \sqrt{\frac{\mathcal A_P + \mathcal A_{NP}}{\mathcal A_P + \mathcal A_{NP} + \mathcal A_P\mathcal A_{NP}}} \\
	Q &= \frac{\mathcal A_{NP}}{\mathcal A_P + \mathcal A_{NP} + \mathcal A_P\mathcal A_{NP}} .
\end{align}
This model is much simpler as there is only one phase: the polar field always orders the nematic field. For the remainder of this paper, we focus on the nematopolar model where the nematic field prefers to be oriented and the polar field prefers to be isotropic.\\

\section{Critical exponents}
\label{app:crit_exps}

For a magnetization $m$, the critical exponents $\alpha$, $\beta$, and $\gamma$ are defined by specific heat $C \equiv -T \p{^2\mathcal F}{T^2}\propto (T - T^*)^{-\alpha}$, magnetization $m \propto (T - T^*)^{\beta}$, and susceptibility $\chi \equiv \left.\p{m}{h}\right|_{h=0} \propto (T - T^*)^{-\gamma}$, respectively, where $h$ is external field. For the N-I line, we assume $2 - \mathcal A_{NP}  \approx T^* - T$, and for the N-NP curve, we assume $\mathcal A_P^* - \mathcal A_P \approx T - T^*$. For the N-I line, the magnetization $m = Q$, and thus it is just the usual mean-field Ising model, so $\alpha = 0$, $\beta = 1/2$, and $\gamma = 1$.

Recall that the free energy $\mathcal F$ is

\beq \mathcal F = \mathcal A_{NP} (Q - p^2)^2 + \mathcal A_P p^2 + (1-Q^2)^2 .\eeq

We first find the critical exponents for the N-I line. Here, $Q$ is the magnetization, and $p=0$. The free energy becomes

\beq \mathcal F = \mathcal A_{NP} Q^2 + (1-Q^2)^2 .\eeq
where we assume $\mathcal A_{NP}^* - \mathcal A_{NP} \approx T^* - T$. This is simply the usual mean-field Ising model with the associated critical exponents.

For the N-NP curve, we have two different analogs of magnetizations: $p$, and the locking order parameter $\sigma \equiv Q - p^2 - \frac{\mathcal A_P}{2\mathcal A_{NP}}$. We show here that generically, the critical exponents for $p$ are $\alpha = 0$, $\beta = 1/2$, and $\gamma = 1$, and at the tricritical point $(\mathcal A_P, \mathcal A_{NP}) = (8/\sqrt{27},4/3)$, the critical exponents for $p$ shift to $\alpha = 1$, $\beta = 1/4$, and $\gamma = 1/2$. For $\sigma$, the critical exponents are $\alpha = 0$, $\beta = 1$, and $\gamma = 0$. 

Recall that in the nematic phase, we have
\beq p = 0, \qquad Q = \sqrt{1 - \mathcal A_{NP}/2} \eeq
and in the nematopolar phase, we have
\beq p = \sqrt{Q - \frac{\mathcal A_P}{2\mathcal A_{NP}}}, \qquad Q = Q(\mathcal A_P) \neq 0 \eeq
We now find the critical exponents for the N-NP curve. Here, we have two choices of magnetization: $p$, or $\sigma$. We first consider $p$ for magnetization. Then the free energy becomes
\beq \mathcal F = \mathcal A_{NP} (Q - p^2)^2 + \mathcal A_P p^2 + (1-Q^2)^2 - hp ,\eeq
where we assume $\mathcal A_P^* - \mathcal A_P \approx T - T^*$ and we coupled $p$ to external field $h$. %, and we have used the fact that in the locked phase, $Q = p^2 + \frac{\mathcal A_P}{2\mathcal A_{NP}}$. 

We first consider a generic point on N-NP curve (excluding the tricritical point). We find $\beta$ by recalling that in the nematopolar phase, since $Q$ is independent of $\mathcal A_{NP}$, then
\beq p = \sqrt{Q - \frac{\mathcal A_P}{2\mathcal A_{NP}}} \sim \sqrt{\mathcal A_P^* - \mathcal A_P} \eeq
giving $\beta = 1/2$. By inspection, it follows that $T\p{^2\mathcal F}{T^2} \approx 1$, giving $\alpha = 0$. The hyperscaling relation $\alpha + 2\beta + \gamma = 1$ implies that $\gamma = 1$.

We now evaluate the critical exponents at the tricritical point. Since at the tricritical point, $\p{^2\mathcal F}{Q^2} = 0$, we also have $Q - Q_c \approx \sqrt{\mathcal A_P^* - \mathcal A_P}$, which from $Q = \sqrt{p^2 - \frac{\mathcal A_P}{2\mathcal A_{NP}}}$ leads to $p \approx (\mathcal A_P^* - \mathcal A_P)^{1/4}$, and thus $\beta = 1/4$. Also, by the chain rule, we can write
\beq \p{^2\mathcal F}{\mathcal A_P^2} = \p{Q}{\mathcal A_P} \p{^2\mathcal F}{Q^2} \eeq
where we have used the fact that $\p{\mathcal F}{Q} = 0$. Thus since $Q - Q_c \approx \sqrt{\mathcal A_P^* - \mathcal A_P}$ and $\p{^2\mathcal F}{Q^2} \approx 1$, we find that
\beq  \p{^2\mathcal F}{\mathcal A_P^2} \approx (\mathcal A_P^* - \mathcal A_P)^{-1}\eeq
giving $\alpha = 1$. The hyperscaling relation $\alpha + 2\beta + \gamma = 1$ implies that $\gamma = 1/2$.

We now repeat this exercise for $\sigma$. We first note that $\beta = 1$, since in the nematic phase, $\sigma = Q - \frac{\mathcal A_P}{2\mathcal A_{NP}}$, where $Q$ is independent of $\mathcal A_{NP}$.

In the nematic phase, since $p = 0$, the free energy $\mathcal F$ becomes
\beq F = \mathcal A_{NP} Q^2 + (Q^2 - 1)^2, \eeq
and since $Q$ is independent of $\mathcal A_P$, then $\mathcal F$ is as well, leading to $\alpha = 0$. The hyperscaling relation $\alpha + 2\beta + \gamma = 1$ implies that $\gamma = 0$.

\section{External fields}
\label{app:ext_fields}

We consider the following free energy:
\beq \mathcal F = \mathcal A_{NP} \left|\overleftrightarrow{Q} - \overleftrightarrow{P}\right|^2 + \mathcal A_P|p|^2 + (|Q|^2-1)^2 -\tr \left[\overleftrightarrow{H}\overleftrightarrow{Q}\right] - \vec h \cdot \vec p .\eeq

We consider two cases here in order: (i) $\vec h = 0, \overleftrightarrow{H} \neq 0$, and (ii) $\vec h \neq 0, \overleftrightarrow{H} = 0$.

\subsection{$\vec h = 0$ and $\overleftrightarrow{H} \neq 0$}

Here we analyze the case of $\vec h = 0$ and $\overleftrightarrow{H} \neq 0$. When $\vec h= 0$, the free energy $\mathcal F$ becomes
\beq \mathcal F = \mathcal A_{NP} \left|\overleftrightarrow{Q} - \overleftrightarrow{P}\right|^2 + \mathcal A_P|p|^2 + (|Q|^2-1)^2 -\tr \left[\overleftrightarrow{H}\overleftrightarrow{Q}\right] \eeq
which is minimized when $\overleftrightarrow{H}$, $\overleftrightarrow{Q}$, and $\overleftrightarrow{P}$ are all aligned. Without loss of generality, we choose coordinates $x$ and $y$ such that $H^{xy} = 0$, and thus by suitable choices of rotations, we can assume $Q^{xy} = 0$ and that $p^y = 0$. This leads to
\beq \mathcal F = \mathcal A_{NP} (Q - p^2)^2 + \mathcal A_P p^2 + (Q^2-1)^2 -2 H Q \eeq
where now $Q, p, H$ denote the magnitudes of $Q, p, H$. Thus in equilibrium,
\begin{subequations}
\begin{align}
	\p{\mathcal F}{p} &= -4\mathcal A_{NP} p (Q - p^2) + 2\mathcal A_P p = 0 \\
	\p{\mathcal F}{Q} &= 2\mathcal A_{NP} (Q - p^2) + 4 Q(|Q|^2 - 1) - 2 H = 0
\end{align}
	\label{eq:equil_external_nonzero_H}
\end{subequations}
%By inspection, Eq.~\eqref{eq:F_ext} (upon setting $h=0$) is minimized when $H$, $Q$, and $pp$ are all aligned, i.e., that they have the same phase. Without loss of generality and for notational convenience, for the rest of this section $Q$, $p$, and $H$ will now denote the magnitudes of $Q$ $p$, and $H$, respectively. (For convenience, since Eq.~\eqref{eq:F_ext} is even in $p$, we are choosing $p\ge 0$). In this case, Eqs.~\eqref{eq:equil_external} simplify to
From Eq.~\eqref{eq:equil_external_nonzero_H}(a), two possible solutions are $p=0$ and $p\neq 0$. We first consider the case $p=0$.

\subsubsection{$p=0$}
We first consider the case $p=0$. When $p=0$, Eq.~\eqref{eq:equil_external_nonzero_H}(b) reduces to
\beq \p{\mathcal F}{Q} = \mathcal A_{NP} Q + 2 Q(Q^2 - 1) - H = 0 %\label{eq:equil_external_nonzero_H} 
\eeq

%From matching with Eq.~\eqref{eq:cubic_depressed}, 
%\beq a = \frac{\mathcal A_{NP}}{2}-1, \qquad b =  - \frac{H}{2} \eeq
%Since $H > 0$, then $b < 0$, which means that the only non-negative solution is
Since $H > 0$, the solution of $Q = 0$ is no longer allowed, and thus the N-I phase boundary disappears. In particular, the only non-negative solution is

\beq Q = \sqrt[3]{\frac{H}{4} + \sqrt{\frac{1}{27}\left(\frac{\mathcal A_{NP}}{2}-1\right)^3 + \frac{H^2}{16}}} + \sqrt[3]{\frac{H}{4} - \sqrt{\frac{1}{27}\left(\frac{\mathcal A_{NP}}{2}-1\right)^3 + \frac{H^2}{16}}}\eeq

\subsubsection{$p \neq 0$}

We now assume $p\neq0$. Then Eq.~\eqref{eq:equil_external_nonzero_H}(a) simplifies to
\beq Q - p^2 = \frac{\mathcal A_P}{2\mathcal A_{NP}}\eeq
and upon substitution of the above into Eq.~\eqref{eq:equil_external_nonzero_H}(b), we learn that
\beq \frac{\mathcal A_P}{2} + 2 Q(Q^2 - 1) - H = 0 \eeq

For sufficiently small $H$, i.e., $H \le \mathcal A_P/2$, the solution is
\beq Q = 2 \, \re{\sqrt[3]{-\frac{\mathcal A_P-2H}{8} + i\sqrt{|D|}}} \eeq
with $D = -\frac{1}{27} + \frac{1}{64}\left(\mathcal A_P - 2H\right)^2$, provided $\mathcal A_P - 2H \le 8/\sqrt{27}$ and $Q \ge \mathcal A_P/(2\mathcal A_{NP})$.

We now show that the N-NP phase boundary is deformed, but remains a continuous phase transition. We find the boundary by setting $p=0$, $Q = \mathcal A_P/(2\mathcal A_{NP})$, and finding the region of parameter space where these conditions and Eq.~\eqref{eq:equil_external_nonzero_H} are satisfied. Doing so gives
\beq \frac{\mathcal A_P}{2} + \frac{\mathcal A_P}{\mathcal A_{NP}}\left(\frac{\mathcal A_P^2}{4\mathcal A_{NP}^2} - 1\right) = H .\eeq

\subsection{$\vec h \neq 0$ and $\overleftrightarrow{H} = 0$}

Here we analyze the case of $\vec h \neq 0$ and $\overleftrightarrow{H} = 0$. When $\overleftrightarrow{H}=0$, the free energy $\mathcal F$ reduces to
\beq \mathcal F = \mathcal A_{NP} |\overleftrightarrow{Q} - \overleftrightarrow{P}|^2 + \mathcal A_P|p|^2 + (|Q|^2-1)^2 - \vec h \cdot \vec p ,\eeq
which is minimized when $\vec p$ is aligned with $\vec h$, and $\overleftrightarrow{Q}$ is aligned with $\overleftrightarrow{P}$. Without loss of generality, we choose $\vec h$ to point in the $\hat x$ direction. $p$ then points in the $\hat x$ direction as well, and $Q^{xy} = 0$. This leads to at equilibrium,
\begin{subequations}
	\begin{align}
		\p{\mathcal F}{p} &= -2\mathcal A_{NP} p (Q - p^2) + \mathcal A_P p - h = 0 \\
		\p{\mathcal F}{Q} &= \mathcal A_{NP} (Q - p^2) + 2 Q(Q^2 - 1) = 0
	\end{align}
	\label{eq:equil_external_nonzero_h}
\end{subequations}
\noindent From Eq.~\eqref{eq:equil_external_nonzero_h}(a), we see that $p = 0$ is not a solution, and hence from Eq.~\eqref{eq:equil_external_nonzero_h}(b), we see that $Q = 0$ is also not a solution. Hence, for non-zero $h$, the nematic and isotropic phases no longer exist -- only the nematopolar phase exists. %Now the question becomes, are there multiple locked phases, or is it unique?
By continuity, for small deformation $h$, the two continuous phase transitions disappear.%, while the 1st order phase transitions persist.

\section{Numerical details}
\label{app:numerics}
\begin{figure}
	\centering
	\includegraphics[width=.85\columnwidth]{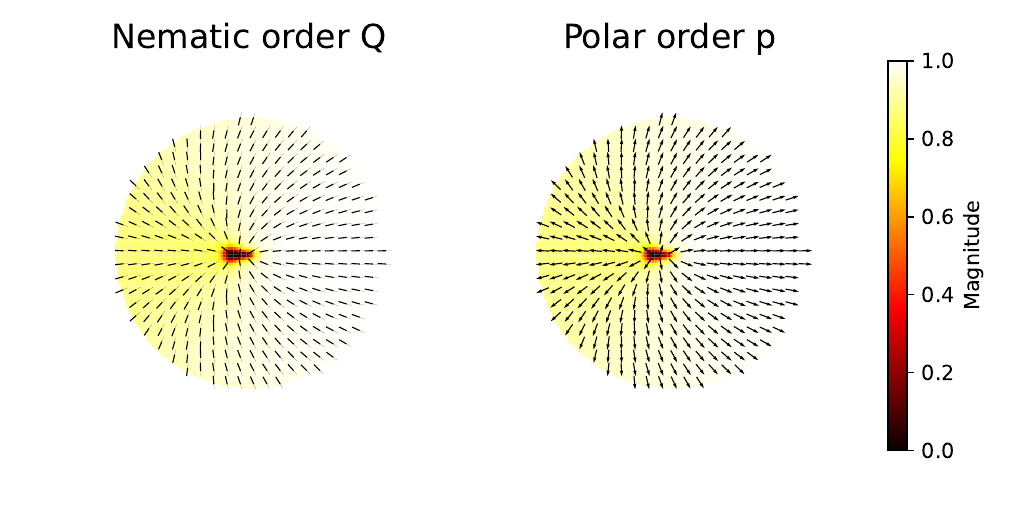}\\
    \includegraphics[width=.85\columnwidth]{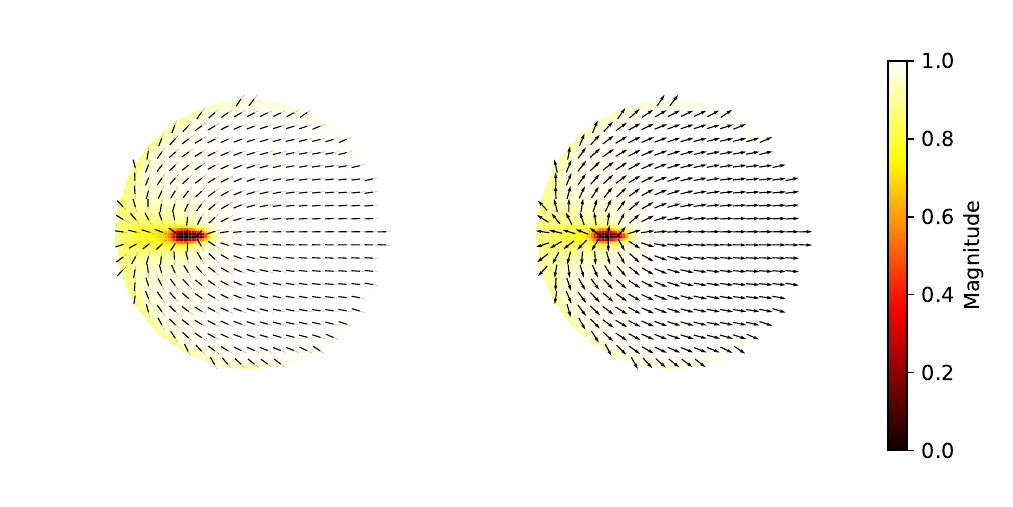}\\
	\includegraphics[width=.85\columnwidth]{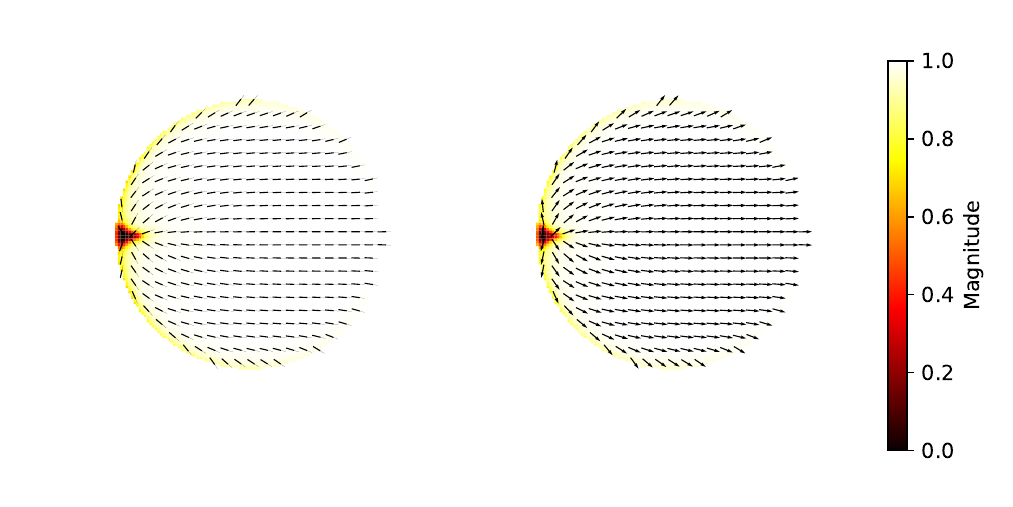}
	\caption{Key snapshots of simulations of plots of $p$ and $Q$ in tightly locked regime and large external field $h$, where color denotes the magnitudes. Snapshots in each column are time-ordered from top to bottom. Initially, the topological defects are at the origin, but they are eventually expelled from the bulk and absorbed by the boundary. %The $+1$ polar topological defect is expelled from the bulk. 
    Parameters used: $\mathcal A_P = 1, \mathcal A_{NP} = 4, h = 4$.}
	\label{fig:expel}
\end{figure}

In our numerical simulations, we minimize the free energy by relaxational dynamics. We do so numerically using the method of lines~\cite{schiesser2012numerical}, where the temporal evolution is performed through a predictor-corrector scheme~\cite{press1992multistep} and spatial derivatives are evaluated using five-point stencil central differences. In the absence of an external field, this method led to steady-states, whereas in the presence of an external field, the topological defects moved until they reachedd the boundary, see Fig.~\ref{fig:expel} for representative snapshots of movement of the topological defects.

\end{document}